\documentclass[%
aps,
prb,%
amsmath,
amssymb,
reprint,%
citeautoscript,
notitlepage,
showpacs,
superscriptaddress
]{revtex4-1}

\usepackage{subfigure}
\usepackage{pdfsync}
\usepackage{color}
\usepackage{graphicx}
\usepackage{dcolumn}
\usepackage{bm}
\usepackage{xifthen}

\usepackage{longtable}
\usepackage{multirow}

\usepackage{eso-pic}
\usepackage{fix-cm}

\usepackage{silence}
\definecolor{dark-blue}{rgb}{0,0,1}
\usepackage[%
linkcolor=dark-blue,
citecolor=dark-blue,
urlcolor=dark-blue,
colorlinks=true,
]{hyperref}

\newboolean{eV}
  \setboolean{eV}{true}

\newboolean{ext_bbl}
  \setboolean{ext_bbl}{true}

\newcommand{\sss}{\scriptscriptstyle}
\newcommand{\sst}{\scriptstyle}

\newcommand{\stext}[1]{\sss \text{#1} \sst}

\newcommand{\eps}{\ensuremath{\text{\large $\boldsymbol{\varepsilon}$}}}
\newcommand{\GMA}{\ensuremath{\text{\large $\boldsymbol{\gamma}$}}}

\makeatletter
  \AddToShipoutPicture{%
	\setlength{\@tempdimb}{.55\paperwidth}%
	\setlength{\@tempdimc}{.5\paperheight}%
	\setlength{\unitlength}{1pt}%
	 \put(\strip@pt\@tempdimb,\strip@pt\@tempdimc){\makebox(0,0){\rotatebox{45}{\textcolor[gray]{0.9}{\fontsize{2cm}{2cm}\textsf{PRB manuscript}}}}}}

\makeatother

\begin{document}

\title{Landau Level optical Hall effect spectroscopy on two- and three-dimensional layered materials with graphene and graphite as examples}

\author{P.~K\"{u}hne}
		\email{kuehne@ifm.liu.se}
		\affiliation{Terahertz Materials Analysis Center, Department of Physics, Chemistry and Biology, IFM, Link\"{o}ping University, SE-58183 Link\"{o}ping, Sweden}
		\homepage{https://www.ifm.liu.se/materialphysics/semicond/}

\author{V.~Darakchieva}
		\affiliation{Terahertz Materials Analysis Center, Department of Physics, Chemistry and Biology, IFM, Link\"{o}ping University, SE-58183 Link\"{o}ping, Sweden}

\author{M.~Schubert}
		\affiliation{Terahertz Materials Analysis Center, Department of Physics, Chemistry and Biology, IFM, Link\"{o}ping University, SE-58183 Link\"{o}ping, Sweden}
		\affiliation{Department of Electrical and Computer Engineering and Center for Nanohybrid Functional Materials, University of Nebraska-Lincoln, Lincoln, Nebraska 68588, USA}
		\affiliation{Leibniz-Institut f\"{u}r Polymerforschung Dresden e.V., Dresden, 01069, Germany}

\date{\today}

\begin{abstract}
We present a comprehensive study of the band structure of two- and three-dimensional hexagonal layered materials using Landau Level optical Hall effect spectroscopy investigations, employing graphene and graphite as model systems. We study inter-Landau-level transitions in highly oriented pyrolytic graphite and a stack of multilayer graphene on C-face 6\textit{H}-SiC, using data from reflection-type optical Hall effect measurements in the mid-infrared spectral range at sample temperatures of $T=1.5$~K and magnetic fields up to $B=8$~T. We describe a comprehensive dielectric polarizability model for inter-Landau-level transitions, which permits analysis of the optical Hall effect data. From their magnetic field dependence we identify sets of H- and K-point inter-Landau-level transitions in graphite and sets of inter-Landau-level transition from decoupled graphene single, and coupled graphene bi-, tri-, and quad-layer for the multilayer graphene stack. For inter-Landau-level transitions in decoupled graphene single-layers and H-point transitions in graphite we observe polarization mode preserving behavior, requiring symmetric magneto-optical contributions to the corresponding dielectric tensors. Inter-Landau-level transitions in coupled graphene layers as well as K-point transitions in graphite exhibit polarization mode mixing behavior, requiring antisymmetric magneto-optical contributions to the corresponding dielectric tensors. From the circular-polarization-averaged inter-Landau-level transition energies and the energy splitting between right- and left-handed circular polarized inter-Landau-level transitions we determine the model parameters in the Slonczewski-Weiss-McClure band structure approximation for two- and three-dimensional hexagonal layered materials, with graphene and graphite as examples.
\end{abstract}



\maketitle

\section{Introduction}\label{sec:intro}
Since the discovery of graphene,\cite{NovoselovS306_2004} the first two-dimensional topological material (2D material), the number of known 2D materials is growing, and has come to include other elemental 2D materials such as silicene\cite{doi:10.1063/1.3419932,PhysRevLett.108.155501} and germanene\cite{Dvila2014}, compound 2D materials, such as boron nitride\cite{ADFM:ADFM201504606}, and tungsten diselenide\cite{Coleman568}, transition metal dichalcogenide monolayers such as molybdenite\cite{Li201533} and 2D transition metal carbides, MXenes\cite{ADMA:ADMA201102306}. In general, 2D materials posses novel electronic and optical properties that are significantly different from the properties of their three-dimensional parent bulk materials enabling potential applications in, e.g., electronic devices, fiber lasers\cite{Sobon2016,C5NR06981E}, photovoltaics\cite{ding:15,C4CS00455H}, batteries\cite{AENM:AENM201600025,AENM:AENM201600671} and water purification systems\cite{C6NR04508A}. The different properties of three- and two-dimensional layered materials are reflected in their corresponding band structures. A precise knowledge of the band structure is a key to full understanding of the electronic and optoelectronic properties. Experimental access to band structure parameters is crucial for tuning theoretical understanding and for design, for example, of electronic and optoelectronic device structures. Optical methods for experimental characterization are highly desirable because of their non-destructive and non-invasive nature.

Excellent model systems to study similarities and differences between band structures of two- and three-dimensional layered hexagonal materials are $\alpha$-graphite and graphene. Graphite consists of individual, Van der Waals bonded graphene layers, where layer stacking results in a hexagonal structure. The distance between the layers is 0.335~nm.\cite{delhaes2000graphite} Graphene consists of two-dimensional, covalently bound carbon atoms  arranged in a honeycomb lattice with 0.142~nm separation.\cite{delhaes2000graphite} Each atom only satisfies three bonds. The remaining electrons render both graphene and graphite electrically conductive, where conductivity  is large parallel to the graphene planes and low perpendicular to the planes. 

Graphene with its unprecedented properties such as its ultra high free charge carrier mobility (hundreds of times larger than in silicon)\cite{Banszeruse1500222}, has the potential to make a profound impact in information and communication technology in the short and long term. Integrating graphene components with silicon-based electronics allows not only substantial performance improvements but, more importantly, enables completely new applications. By exploiting the unique electrical and optical properties of graphene, novel electronics systems with ultra-high speed of operation and electronic devices with transparent and flexible form factors can be developed. Graphene materials also allow to combine structural functions with embedded electronics in an environmentally sustainable and biocompatible manner\cite{Pinto2013188}. Graphite has a variety of applications, e.g.,  as a promising anode material in advanced Li-ion batteries\cite{ESE3:ESE395}, as functional refractories for casting of steel\cite{IJAC:IJAC02174}, for the manufacturing of crucibles\cite{Chernyavets2008}, as carbon brushes in electrical motors\cite{toliyat2004handbook}, in nuclear industry as moderating rods\cite{stacey2007nuclear}, as engineering material in aviation industries\cite{20145}, as lubricant\cite{bruce2010crc}, and paint and coatings\cite{Azim20061}.

The study of the fundamental physical properties of graphite is essential for understanding the properties of new nanostructured sp$^2$ carbon forms, such as graphenes, fullerenes and carbon nanotubes. Many physical properties of graphite and graphene are conveniently described by the Slonczewski-Weiss-McClure~(SWC) model,\cite{PhysRev.109.272,PhysRev.108.612} a band model with seven tight-binding model parameters for the Brillouin zone around the H-K-H axis.
A suitable method to access the band structure parameter of the SWC model is Landau level spectroscopy, which has been widely applied for graphite\cite{DRESSELHAUS1966465,doi:10.1143/JPSJ.33.1619,Toy77,PhysRevB.19.4224,PhysRevB.21.827,doi:10.1143/JPSJ.47.199,doi:10.1080/00018730110113644} and graphene.\cite{SadowskiPRL97_2006,OrlitaPRL101_2008,HenriksenPRL100_2008,OrlitaPRL102_2009,OrlitaPRL107_2011,OrlitaPRB83_2011,OrlitaPRL107_2011,PhysRevB.80.165406,PhysRevB.78.235408,PhysRevB.76.201401,PhysRevB.80.165406,PhysRevB.85.195406} Under the influence of a magnetic field the momentum of free charge carriers is subject to the Lorentz force. If the scattering time is high enough cyclotron orbits of electrons (holes) in two-dimensional confinement can become quantized (Landau level; LL), resulting in a discrete energy spectrum. Landau quantization can appear in two dimensional and quasi two dimensional materials, for example, in decoupled graphene layers, coupled graphene layers, 2D electron gases, or graphite. Landau levels in two- and three-dimensional materials differ, in general. Technological advances have made preparation of two- and three-dimensional layered hexagonal materials possible, where for example few ($m$) layers of graphene can be grown on SiC\cite{doi:10.1063/1.4908216,Kuehne14}. An interesting question is the evolution of LL levels and allowed electronic transitions as layered hexagonal materials transform from pure 2D to de-facto infinite 3D crystal structures, such as from graphene to graphite. In addition to its dependence on the electronic level structure, LL transitions also depend strongly on polarization. Few polarization-resolved LL spectroscopy experiments have been been reported.\cite{doi:10.1143/JPSJ.33.1619,Toy77,PhysRevB.19.4224,PhysRevB.21.827,PhysRevB.85.195406} A systematic study of polarization-resolved LL spectroscopy can be performed by the optical Hall effect (OHE).\cite{Schubert:16} The OHE is a physical phenomenon, which describes the occurrence of transverse and longitudinal magnetic field-induced birefringence, caused by the nonreciprocal,~\cite{Weiglhofer03} magneto-optical response of electric charge carriers. The OHE grants access to the magneto-optical dielectric tensor\cite{SchubertJOSAA20_2003} and thereby to the polarization selection rules,\cite{KuehnePRL111_2013} that is, whether individual transitions are polarization preserving or polarization mixing.

In this paper, we employ our mid-infrared sub-system of the integrated mid-infrared, far-infrared, and terahertz OHE instrument\cite{KuehneRSI2014} to study and compare inter-LL transitions in highly oriented pyrolytic graphite and multilayer graphene on 6\textit{H}-SiC for sample temperatures of $T=1.5$~K and magnetic fields up to $B_{\perp}=8/\sqrt{2}$~T. We show that using our comprehensive polarization-resolving approach, effects from differing absorptions of right- and left-handed polarized light can be distinguished. We determine the polarization selection rules for different types of inter-LL transitions. We use the polarization selection rules, together with first-principle considerations, to derive the symmetry properties of the underlying dielectric tensor. Further, from the magnetic field dependence, we identify the location of inter-LL transitions in the Brillouin zone, the stacking and coupling of the individual sheets in multilayer graphene. Finally, we apply the dielectric tensor model to access the SWC tight-binding parameter of graphite and epitaxial graphene.

\section{Theory}\label{sec:theory}
\subsection{Magneto-optical dielectric tensors}\label{sec:DFs}
The optical response of a material with arbitrary anisotropy can be uniquely described by the dielectric tensor $\eps$. The dielectric tensor is a second-rank tensor, which connects the electric field vector $\mathbf{E}$ with the electric displacement field vector $\mathbf{D}$. The electric displacement field describes the electric flux density at the surface of a medium, and can be written as
\begin{equation}
\label{eqn:def-D-field}
	\mathbf{D}
	=
	\varepsilon_0\mathbf{E}+\mathbf{P}
	=
	\varepsilon_0\mathbf{E}+\varepsilon_0\bm{\chi}\mathbf{E}
	=
	\varepsilon_0\left(\mathbf{I}+\bm{\chi}\right)\mathbf{E}
	=
	\varepsilon_0\eps\mathbf{E}\;,
\end{equation}
where $\varepsilon_0$, $\mathbf{P}$, $\mathbf{I}$, and $\bm{\chi}$ denote the electric vacuum permittivity, the electric polarization vector, the $3\times 3$ unit matrix, and the electric susceptibility tensor of the medium, respectively. For a linear optical material response, the total dielectric tensor may be written as the sum of electric susceptibility tensors
\begin{equation}
	\label{eqn:sum-suscepti}
	\eps
	=
	\mathbf{I}+\bm{\chi}
	=
	\mathbf{I}+\sum_k \bm{\chi}_{_{k}}\;,
\end{equation}
where each $\bm{\chi}_{_{k}}$ is a second-rank tensor related to an independent mechanism of polarization within the medium, such as phonon modes, electronic transitions, or magneto-optical effects.

If the dielectric tensor of a material without a magnetic field is given by $\eps_{_{\hspace{-1pt}\mathbf{B}=0}}=\mathbf{I}+\bm{\chi}_{_{\hspace{-1pt}\mathbf{B}=0}}$, the optical response induced by a given quasi static magnetic field $\mathbf{B}$ is then expressed by the electric susceptibility tensor $\bm{\chi}_{_{\hspace{-1pt}\mathbf{B}}}$. The magneto-optical dielectric tensor describing the OHE can be written as
\begin{equation}\label{eqn:MO-DF}
		\eps_{_{\hspace{-1pt}\mathbf{B}}}
		=
		\mathbf{I}
		+
		\bm{\chi}_{_{\hspace{-1pt}\mathbf{B}=0}}
		+\;
		\bm{\chi}_{_{\hspace{-1pt}\mathbf{B}}}\;.
\end{equation}

The magneto-optical response typically originates from bound and unbound charge carriers subject to the Lorentz force. Thus the magneto-optical response is optically anisotropic and non-reciprocal in time, i.e., the magnetic field induced electric susceptibility tensor is not symmetric $\bm{\chi}_{_{\hspace{-1pt}\mathbf{B}}}\neq\bm{\chi}_{_{\hspace{-1pt}\mathbf{B}}}^{\text{T}}$, where the superscript $\text{T}$ stands for transposed.\cite{SchubertIRSEBook_2004,weiglhofer2003introduction} This is reflected by different electric susceptibilities for right- and left-handed circularly polarized light, $\chi_{\mathrm{+}}$ and $\chi_{\mathrm{-}}$, respectively.\cite{SchubertJOSAA20_2003, Hofmannpss205_2008} Without loss of generality, if the magnetic field $\mathbf{B}$ is pointing in \textit{z}-direction, the polarization vector \mbox{$\mathbf{P}=\varepsilon_0\bm{\chi}\mathbf{E}$} can be described by arranging the electric fields in their circularly polarized eigensystem \mbox{$\mathbf{E}_e = (E_x + \mathrm{i}E_y, E_x - \mathrm{i}E_y, E_z) = (E_{+}, E_{-}, E_z)$} by $\mathbf{P}_e = \varepsilon_0\bm{\chi}_e \mathbf{E}_e = \varepsilon_0(\chi_{\mathrm{+}}E_{+}, \chi_{\mathrm{-}}E_{-}, 0)$, where $\text{i}=\sqrt{-1}$ is the imaginary unit.\cite{SchubertADP15_2006,HofmannRSI77_2006} Transforming $\mathbf{P}_e$ back into the laboratory system the magnetic field induced electric susceptibility tensor takes the form:\cite{SchubertADP15_2006, HofmannRSI77_2006} 
\begin{equation} 
	\bm{\chi}_{_{\hspace{-1pt}\mathbf{B}}}
	=
	\frac{1}{2}
	\begin{pmatrix}
	\hspace{11pt}(\chi_{\textbf{\tiny{+}}}+\chi_{\textbf{\tiny{--}}}) & \text{i}(\chi_{\textbf{\tiny{+}}}-\chi_{\textbf{\tiny{--}}}) & 0 \\
	   -\text{i}(\chi_{\textbf{\tiny{+}}}-\chi_{\textbf{\tiny{--}}}) & \hspace{3pt}(\chi_{\textbf{\tiny{+}}}+\chi_{\textbf{\tiny{--}}}) & 0 \\
			0																						& 0																						& 0 \\
	\end{pmatrix}\;.
	\label{eqn:general_susceptibility}
\end{equation}
Note, under spacial field inversion, $\mathbf{B}\rightarrow -\mathbf{B}$, the electric susceptibilities for left- and right-handed circularly polarized light interchange, $\chi_{\mathrm{+}}\rightleftharpoons \chi_{\mathrm{-}}$. Hence, $\bm{\chi}_{_{\hspace{-1pt}\mathbf{B}}}$ is only diagonal if $\chi_{\mathrm{+}}=\chi_{\mathrm{-}}$, and is non-diagonal otherwise, with anti-symmetric off-diagonal elements.

\subsubsection{Lorentz-Drude model}
In the correspondence limit, electric charge carriers subject to a quasi static magnetic field $\mathbf{B}$ obey Newton's equation of motion
	\begin{equation}
	\mathbf{m}\mathbf{\ddot{x}}
	+
	\mathbf{m}\GMA\mathbf{\dot{x}}
	+
	\mathbf{m} \omega_0^2 \mathbf{x}
	=
	q \mathbf{E} + q (\mathbf{\dot{x}} \times \mathbf{B})\;,
	\label{eqn:fcc-newton}
	\end{equation}
where $\mathbf{m}$, $q$, $\GMA$, $\boldsymbol{\mu}=q\mathbf{m}^{-1}\GMA^{-1}$, $\mathbf{x}$, and $\omega_0$ represent the effective mass tensor, the electric charge, damping constant tensor, the mobility tensor, the spatial coordinate of the charge carrier, and the eigen frequency of the undamped system without external excitation and magnetic field, respectively. For a time harmonic electromagnetic plane wave with an electric field $\mathbf{E}\rightarrow\mathbf{E}\exp(\text{i}\omega t)$ with angular frequency $\omega$, the time derivative of the spacial coordinate of the charge carrier is $\mathbf{\dot{x}}=\mathbf{v}\exp(\text{i}\omega t)$, where $\mathbf{v}$ is the velocity of the charge carrier. With the current density, $\mathbf{j}=n q \mathbf{v}$, Eq.~(\ref{eqn:fcc-newton}) reads
	\begin{equation}
		\mathbf{E}
		=
		\frac{1}{\hat{n}q}
		\left[
			\text{i}
			\frac{\mathbf{m}}{q\omega}
			\left(
				\omega_0^2\mathbf{I}-\omega^2\mathbf{I}-\text{i}\omega\GMA
			\right)
			\mathbf{j}
			+
			(\mathbf{B} \times \mathbf{j})
		\right]\;,
	\end{equation}
where $\hat{n}$ is the charge carrier density. Using \mbox{$\mathbf{E}= \boldsymbol{\sigma}^{-1} \mathbf{j}$} and \mbox{$\eps=\bm{I}+\frac{1}{\text{i} \varepsilon_0 \omega}\boldsymbol{\sigma}$} with the conductivity tensor $\boldsymbol{\sigma}$, the electric susceptibility for charge carriers subject to the external magnetic field $\mathbf{B}$ can be expressed as (Lorentz-Drude model)\cite{Yu99}
\begin{equation}\label{eqn:fulldrude}
	\chi_{ik}
	=
	\frac{\hat{n}q^2}{\varepsilon_0}
		\left[
			m_{ik} (\omega_0^2-\omega^2-\text{i}\omega\gamma_{ik}) 
			-
			\text{i}\omega\epsilon_{ijk} q B_j
		\right]^{-1}\;,
\end{equation}
where $\epsilon_{ijk}$ is the Levi-Cevita-Symbol.
\ifthenelse{\boolean{ext_bbl}}{\cite{Note_PRB1}}{\footnote{In the following equation the Einstein notation is used, and the covariance and contra variance is ignored since all coordinate systems are Cartesian (The summation is only executed over pairs of lower indices).}}

\paragraph{Polar lattice vibrations}\label{sec:PhononDF} For isotopic effective mass tensors the cyclotron frequency $\omega_\text{c}=\frac{q |B|}{m}$ can be defined. For mass values in the order of atomic masses, e.g., in the case of polar lattice vibrations, the cyclotron frequency is several orders of magnitude smaller than for effective electron masses, and can be neglected for magnetic fields and spectral ranges discussed in this paper. Therefore, the dielectric tensor of polar lattice vibrations $\eps^{\text{\tiny{L}}}$ can be approximated using Eq.~(\ref{eqn:fulldrude}) with $\mathbf{B}=0$. When assuming isotropic effective mass and mobility tensors, the result is a simple harmonic oscillator function with Lorentzian-type broadening.\cite{Pidgeon80,Kittel86,Yu99} For materials with orthorhombic and higher symmetry and multiple optical excitable lattice vibrations, the dielectric tensor can be diagonalized to
\begin{equation}
	\eps^{\text{\tiny{L}}}
	=
	\begin{pmatrix}
		\varepsilon_{x}^{\text{\tiny{L}}} & 0	&	0 \\
		0 & \varepsilon_{y}^{\text{\tiny{L}}}	&	0 \\
		0 & 0	&	\varepsilon_{z}^{\text{\tiny{L}}} 
	\end{pmatrix}\;,
	\label{Substrate}
\end{equation}
where $\varepsilon_{\text{\tiny{\textit{k}}}}^{\text{\tiny{L}}}$ ($k=\{x,y,z\}$) is given by\cite{Barker64,KUKHARSKII19731761}
\begin{equation}
	\varepsilon_{\text{\tiny{\textit{k}}}}^{\text{\tiny{L}}}=\varepsilon_{\infty,\text{\tiny{\textit{k}}}}\prod^{l}_{j=1}\frac{\omega^2+\text{i}\omega\gamma_{\stext{LO},\text{\tiny{\textit{k,j}}}}-\omega^2_{\stext{LO},\text{\tiny{\textit{k,j}}}}}{\omega^2+\text{i}\omega\gamma_{\stext{TO},\text{\tiny{\textit{k,j}}}}-\omega^2_{\stext{TO},\text{\tiny{\textit{k,j}}}}}\;,
	\label{eqn:phonon2}
\end{equation}
where $\omega_{\text{LO,\text{\tiny{\textit{k,j}}}}}$, $\gamma_{\text{LO,\text{\tiny{\textit{k,j}}}}}$, $\omega_{\text{TO,\text{\tiny{\textit{k,j}}}}}$, $\gamma_{\text{TO,\text{\tiny{\textit{k,j}}}}}$ and $\varepsilon_{\infty,\text{\tiny{\textit{k}}}}$ denote the \mbox{$k=\{x,y,z\}$} component of the frequency and the broadening parameter of the $j^{\text{th}}$ longitudinal optical (LO), transverse optical (TO) phonon modes and the  high-frequency dielectric constant, respectively, while the index $j$ runs over $l$ modes. Further details can be found in Refs.~\onlinecite{Barker64,Berreman68,Gervais74, HofmannAPL88_2006, HofmannPRB66_2002}, and a detailed discussion on the requirements to the broadening parameters, so $\Im\text{m}\left\{\varepsilon_{\text{\tiny{\textit{k}}}}^{\text{\tiny{L}}}\right\}\geq 0$ is fulfilled, can be found in Ref.~\onlinecite{KasicPRB62_2000}.

\begin{figure*}[htbp]
	\centering
  \includegraphics[
	width=0.98\textwidth]{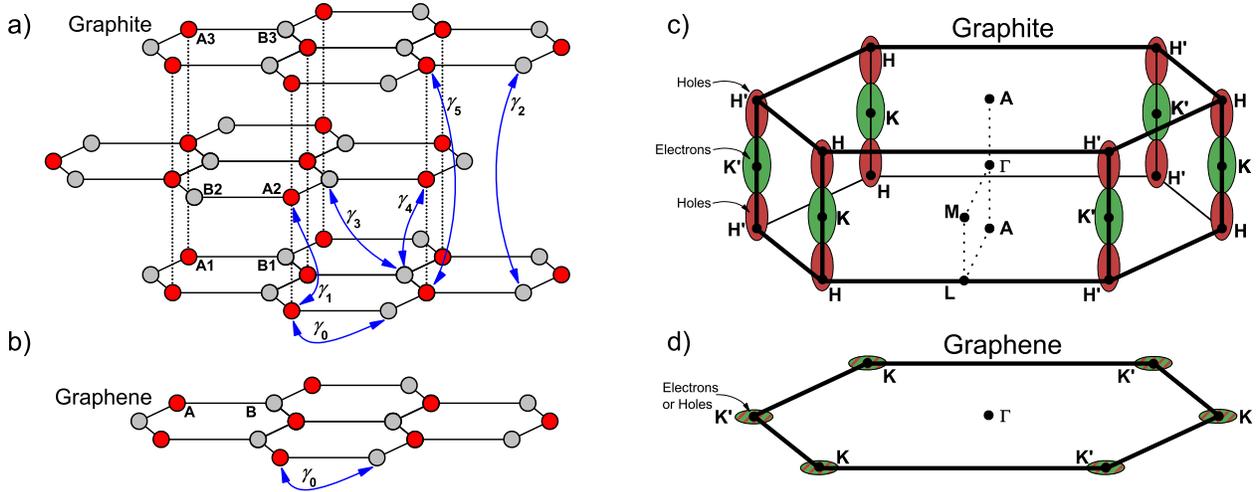}
	\caption{a) Bernal stacked crystal lattice of graphite and b) crystal lattice of graphene. Letters denote the sub-latices (A, B), numbers indicate the layer number (1, 2, 3), and arrows illustrate hopping parameters of the Slonczewski-Weiss-McClure model. c) and d) display the Brillouin zones of graphite and graphene, respectively. Letters indicate the high symmetry points.}
	\label{fig:Crystal_structure}
\end{figure*}

\paragraph{Free charges carriers}\label{sec:DrudeDF}
For free charge carriers no restoring force is present and the eigen frequency of the system is $\omega_0=0$. For isotropic effective mass and conductivity tensors, and for magnetic fields aligned along the \textit{z}-axis Eq.~(\ref{eqn:fulldrude}) can be written in the form
$
	\eps^{\text{\tiny{D}}}_{\text{\tiny{OHE}}}(\mathbf{B})
	=
  \mathbf{I}
	+
	\bm{\chi}^{\text{\tiny{D}}}
	=
  \mathbf{I}
	+
	\bm{\chi}_{_{\hspace{-1pt}\mathbf{B}=0}}^{\text{\tiny{D}}}
	+
	\bm{\chi}_{_{\hspace{-1pt}\mathbf{B}}}^{\text{\tiny{D}}}
$
, with the Drude electric susceptibility tensor for $\mathbf{B}=0$
\begin{equation}
	\bm{\chi}_{_{\hspace{-1pt}\mathbf{B}=0}}^{\text{\tiny{D}}}
	=
	-\frac{\omega_{\text{p}}^2}{\omega (\omega+\text{i}\gamma)}
	\mathbf{I}
	=
	\chi^{\text{\tiny{D}}}
	\mathbf{I}\;,
	\label{eqn:iso_drude}
\end{equation}
where $\omega_{\text{p}}=\sqrt{\frac{\hat{n} q^2}{m\varepsilon_0}}$ is the plasma frequency, and $\chi^{\text{\tiny{D}}}$ is the electric susceptibility of the isotropic Drude dielectric function. For isotropic effective mass and conductivity tensors the magneto-optical electric susceptibility $\bm{\chi}_{_{\hspace{-1pt}\mathbf{B}}}^{\text{\tiny{D}}}$ can be expressed through electric susceptibility functions for right- and left-handed circularly polarized light
\begin{equation}
	\chi_{\mathrm{\pm}}
	=
	\frac{\chi^{\text{\tiny{D}}}}{1\mp \frac{\omega+\text{i}\gamma}{\omega_{\text{c}}}}\;,
	\label{eqn:chi_drude_pm}
\end{equation}
where $\omega_{\text{c}}=\frac{q |B|}{m}$ is the isotropic cyclotron frequency.

\subsubsection{Inter-Landau-level transition model}\label{sec:LandauDF}
\paragraph{Selection rules:} Transitions between Landau levels obey optical selection rules, demanding $n'=n\pm 1$
\ifthenelse{\boolean{ext_bbl}}{\cite{Note_PRB2}}{\footnote{This statement stays correct when effects such as trigonal warping of the band structure in graphite are included\cite{doi:10.1143/JPSJ.40.761,PhysRev.140.A401,doi:10.1143/JPSJ.17.808}}}
for transitions from level $n$ to level $n'$,\cite{SadowskiIJMPB_2007,KoshinoPRB77_2008,Toy77} where the $+$ sign corresponds to right-, and the $-$ sign corresponds to left-handed circularly polarized radiation, respectively.\cite{PhysRevB.21.827} In this paper, Landau levels are denoted using the index $n \in \mathbb{N}_0$ for the quantum number, while inter-Landau-level transitions are denoted using the index $N \in \mathbb{N}$ with
\begin{equation}
	2N=n+n'+1\;.
\end{equation}
Therefore, using optical selection rules, the transitions $\text{L}_n^-\rightarrow \text{L}_{n+1}^+$ and $\text{L}_{n+1}^-\rightarrow \text{L}_n^+$ are both labeled with the index $N$. 

\paragraph{Circular polarization transition model:}
At finite temperatures, $T\neq 0$, Fermi's golden rule can be used to describe the contribution of inter-Landau-level transitions to the dielectric displacement, and the magneto-optical dielectric response tensor can be rendered by a sum of Lorentz oscillators. The broadening parameter of a given Lorentz oscillator may then be interpreted as a mean scattering life time parameter of the corresponding transition, $\tau=\frac{1}{\gamma}$. The electric susceptibilities $\chi_{\mathrm{\pm}}$ for right- and left-handed circular polarized light in Eq.~\ref{eqn:general_susceptibility} are then given by
	\begin{equation}
		\chi_{\pm} = 
		\sum_N \frac{A_{\pm,N}}{E^2_{\pm,N}-E^2- \text{i} E \gamma_{\pm,N}}\;,
		\label{eqn:landau}
	\end{equation}
where $A_{\pm,N}$ and $\gamma_{\pm,N}$ are the amplitude and the broadening parameters of the $N^{\text{th}}$ inter-Landau-level transition, respectively, and the the energy of the $N^{\text{th}}$ inter-Landau-level transition is given by
\begin{equation}
	E_{\pm,N}
	=
	\begin{cases}
		E_{+,N}=E_c(n)-E_v(n+1)\\
		E_{-,N}=E_c(n+1)-E_v(n)\;.
	\end{cases}
	\label{eqn:transition_energies}
\end{equation}
Note that all parameters are functions of the magnetic field. Depending on the exact values of all parameters, two different cases can be discerned here:

\subparagraph{Polarization preserving transitions:} In the case when all parameters for left- and right-handed inter Landau level transitions are equal, Eq.~\ref{eqn:general_susceptibility} describes a diagonal tensor. As a result, reflection and transmission type experiments performed on samples whose surfaces are parallel to the plane of two-dimensional confinement will not reveal conversion of polarization, for example from right- to left-handed, or from $parallel$ ($p$) to $senkrecht$ ($s$) with respect to the plane of incidence. We have previously identified such circumstances with polarization conserving inter-Landau-level transitions.\cite{KuehnePRL111_2013}
\subparagraph{Polarization mixing transitions:} In the general case when parameters for right- and left-handed circular polarized light differ, Eq.~\ref{eqn:general_susceptibility} describes a non-diagonal and antisymmetric tensor. As a result, polarization conversion occurs during reflection and transmission type experiments. We have previously identified such circumstances with polarization mixing inter-Landau-level transitions.\cite{KuehnePRL111_2013}
%

\paragraph{Small energy splitting approximation:} For experimental situations when series of inter-Landau-level transitions occur for right- and left-handed polarized light with nearly equal transition amplitude and broadening parameters and slightly different energies, an approximation can be made to simplify Eq.~\ref{eqn:general_susceptibility}. Consider two Lorentz oscillators for $\chi_{\pm}$ with the same amplitude $A_{+,N}=A_{-,N}=\frac{1}{2}A_N$ and broadening $\gamma_{+,N}=\gamma_{-,N}=\gamma_N$ but different transition energies, such that \mbox{$E_{+,N}=E_{N}+\frac{\delta E_{N}}{2}$} and \mbox{$E_{-,N}=E_{N}-\frac{\delta E_{N}}{2}$}, with \mbox{$\delta E_{N} \ll E_{N}$} and $\delta E\approx \gamma_N$. It can be shown that the on- and off-diagonal elements of $\bm{\chi}_{_{\hspace{-1pt}\mathbf{B}}}$ in Eqn.~\ref{eqn:general_susceptibility} can then be approximated by Lorentz oscillator functions with parameters $A_N,E_N$, and $\gamma_N$, such that

\begin{equation}
\begin{aligned} 
	\bm{\chi}^{\stext{(A)}}_{_{\hspace{-1pt}\mathbf{B}}}
		& \approx
		\begin{pmatrix}
			\hspace{8pt}\chi_{_{\text{A}}} & \chi_{_{\text{A}}} & 0 \\
			-\chi_{_{\text{A}}} & \chi_{_{\text{A}}} & 0 \\
			\hspace{8pt} 0						& 0								& 0
		\end{pmatrix}\;,\quad\quad\text{with}\\
	\chi_{_{\text{A}}}
		&=
		\sum_N \frac{A_{N}}{E^2_{N}-E^2-\text{i} E \gamma_{N}}\;.
		\label{eqn:landau_averaged}
\end{aligned} 
\end{equation}
This approximation is convenient for analysis of inter-Landau-level transitions in materials with slightly differing transition energies for right- and left-handed polarized light, and we exploit this approximation in this work. Note that the small energy splitting approximation describes polarization mixing inter-Landau-level transitions. Note further that this approximation determines averaged energy parameters, $E_N$, and is independent on the actual splitting parameter, $\delta E_N$, for as long as the above made considerations for $A_{\pm,N}$, $E_{\pm,N}$, and $\gamma_{\pm,N}$ hold.

\subsection{Slonczewski-Weiss-McClure band model}
In the single particle picture the band structures of graphite and graphene are often described by the Slonczewski-Weiss-McClure (SWM) model.\cite{PhysRev.109.272,PhysRev.108.612} This tight-binding model uses seven parameters to describe the band structure near the H- and K-points: 
$\Delta$ represents the difference between the on-site energies for the two sub-lattices A and B due to their stacking difference and $\gamma_j$ $(j=0\dots 5)$ are hopping energies (see Fig.~\ref{fig:Crystal_structure}). Further details are given in the \hyperref[appendix:SWC-model]{appendix}.



\subsection{Landau levels in hexagonal Carbon layered materials}
\subsubsection{Single and decoupled graphene layers}
\begin{figure*}[htbp]
	\centering
  \includegraphics[
	width=0.98\textwidth]{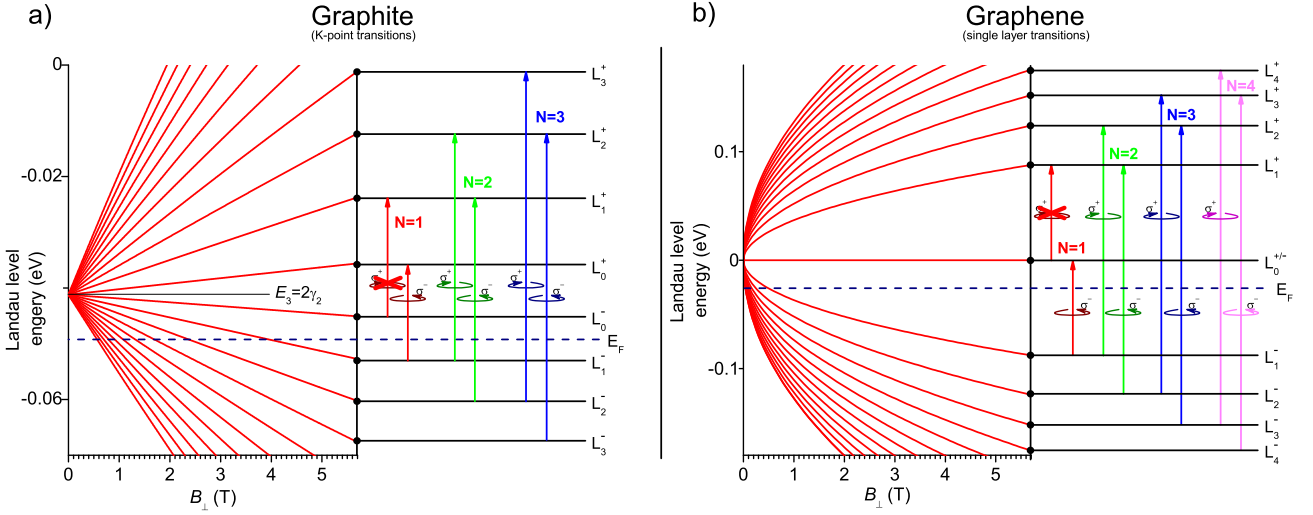}
	\caption{Schema of Landau levels and inter-Landau-level transitions in a) graphite and b) single-layer graphene. In case of graphite the Landau level spectrum is almost linear in $B_{\perp}$ while for single-layer graphene a $\sqrt{B_{\perp}}$ dependence is observed. The arrows indicate allowed transitions between Landau levels with quantum numbers $n$ and $n'$ with $n'=n\pm 1$, where the $+$ and $-$ stand for right- and left-handed circular polarized light, respectively. For a given Fermi-level, indicated by the dashed line, certain transitions are not observed due to the occupation state of the initial and end Landau levels (here: crossed out transitions with $N=1$). Note that we use a simplified notation for the Landau level indices and that for graphite the magnetic field dependence of the levels with a low quantum number is more complex in general.\cite{doi:10.1143/JPSJ.40.761} $B_{\perp}$ is the out-of-plane component of $\mathbf{B}$ with respect to the honeycomb structure of graphite and graphene. The Landau level field dependencies are plotted using parameters from Ref.~\citenum{Chung2002}, where we set $\gamma_3=0$ for graphite and $\Delta=0$ for graphene.}
	\label{fig:Graphene_graphite_principle_LL}
\end{figure*}

In single-layer graphene as well as in multiple, but electronically decoupled graphene layers, electrons or holes are located near the K-point (see Fig.~\ref{fig:Crystal_structure}) and only the nearest neighbor hopping parameter $\gamma_0$ of the SWM-model is relevant. Furthermore, for magnetic fields relevant for this paper the parameter $\Delta$ generates only small offset in the inter-Landau-level transition energies, and is therefore neglected. The eigenvalues of the Hamiltonian for the massless fermions in single-layer graphene~\cite{GeimNM6_2007} (denoted SLG, Fig.~\hyperref[fig:Graphene_graphite_principle_LL]{\ref*{fig:Graphene_graphite_principle_LL} b}) and in decoupled graphene sheets~\cite{OrlitaPRL101_2008} depend on $\sqrt{|B_{\perp}|}$ 
\begin{equation}
	E_{\stackrel{c}{\text{\tiny{\textit{v}}}}}^{\stext{SLG}} =\pm E_0\sqrt{n}\;,
	\label{eqn:SLG_LL}
\end{equation}
with $E_0=v_f\sqrt{2\hbar e|B_{\perp}|}$ and the Fermi velocity \mbox{$v_f=\sqrt{3}\frac{\gamma_0a_0}{2\hbar}$}, where $e$, $\hbar$, $B_{\perp}$, and $a_0$ denote the elementary charge, the reduced Planck constant, the magnetic field perpendicular to the plane of confinement, and the in-plane lattice parameter, respectively. The indicies $c$ and $v$ stand for the conduction and valence band, respectively. Note that the Landau level spectrum (Fig.~\hyperref[fig:Graphene_graphite_principle_LL]{\ref*{fig:Graphene_graphite_principle_LL} b}) of the conduction and valence band is symmetric, i.e., $E_c^{\stext{SLG}}=-E_v^{\stext{SLG}}$. Hence, inter-Landau-level transitions detected in reflection or transmission experiments appear as polarization conserving transitions when amplitude and broadening parameters equal for right- and left-handed circularly polarized light. (See Sect.~\ref{sec:LandauDF}.)

\subsubsection{Coupled graphene layers}
In coupled graphene layers (bi-layer, tri-layer, quad-layer$,\dots,m$-layer) the SWM-model parameters $\gamma_0,\gamma_1,\gamma_3,\gamma_4$ and $\Delta$ are relevant. Here, we use a model from Koshino et.~al.\cite{KoshinoPRB77_2008} for $m$-layer graphene, which neglects $\gamma_3=0,\gamma_4=0$, and $\Delta=0$. Then the eigenvalues of the Hamiltonian of Bernal-stacked $m$-layer graphene can be divided into two branches: a single-layer-like branch and a bi-layer-like branch.\cite{KoshinoPRB77_2008} The single-layer-like transitions only occur in $m$-layer graphene with an odd layer number ($m=3,5,7,\dots$) and exhibit a $\sqrt{B_{\perp}}$-dependence described by Eqn.~\ref{eqn:SLG_LL}. The bi-layer-like transitions (BLG) in $m$-layer graphene follow a sub-linear behavior in $B_{\perp}$\cite{AbergelPRB75_2007,KoshinoPRB77_2008, OrlitaPRL107_2011}
\begin{equation} \label{eqn:LLNLayer}
\begin{split}
	E_{\stackrel{c}{\text{\tiny{\textit{v}}}}}^{\stext{m-BLG}}&
	= \pm \frac{1}{\sqrt{2}}\left[\vphantom{\frac{1}{\sqrt{2}}}(\Gamma_m\gamma_1)^2+\left(2 n+1\right)E_0^2\right. \\
	& \hspace{-6mm} \left. -\sqrt{(\Gamma_m\gamma_1)^4+2\left(2 n+1\right)E_0^2(\Gamma_m\gamma_1)^2+E_0^4}\right]^{1/2},
\end{split}
\end{equation}
where $\Gamma_m
$~\cite{KoshinoPRB77_2008} denotes a parameter, which depends on the number of coupled graphene sheets and determines where in $k_z$-space transitions occur. 

Note that, in principle, energy parameters for left- and right-handed polarized inter-Landau-level transitions for the bi-layer-like branch in $m$-layer graphene differ, but that due to the assumption $\gamma_3=\gamma_4=0$ the Landau level spectrum of the conduction and valence band is symmetric, i.e., $E_c^{\stext{m-BLG}}=-E_v^{\stext{m-BLG}}$, resulting in inter-Landau-level transition energies \textit{independent} of the handedness of the circular polarization. Therefore, in principle, the \textit{circular polarization transition model} (Eqn.~\ref{eqn:landau}) must be applied. However, the splitting between energies for left- and right handed polarization is small. Therefore, the \textit{small energy splitting approximation} with $E_N$ from Eqn.~\ref{eqn:LLNLayer} is used to model experimental observations.

\subsubsection{Graphite}
For graphite all SWM-model parameters are taken into account except $\gamma_3=0$, which causes a trigonal wraping of the band structure and which affects Landau levels with small quantum numbers only.\cite{PhysRev.140.A401,doi:10.1143/JPSJ.40.761} Two sets of inter-Landau-level transitions, corresponding to the H- and K-point (high symmetry points of the Brillouin zone), are allowed.
\ifthenelse{\boolean{ext_bbl}}{\cite{Note_PRB3}}{\footnote{Using Fermi's golden rule it can be shown\cite{PhysRev.140.A401,doi:10.1143/JPSJ.17.808} that the joint density of states (between conduction band and valence band) has singularities at the two high symmetry points of the Brillouin zone, the H- and K-point.}}

For the H-point series the eigenvalues of the Hamiltonian take the same form as for single-layer graphene
\begin{equation}
	E_{\stackrel{c}{\text{\tiny{\textit{v}}}}}^{\stext{H}}
	=
	\pm E_0\sqrt{n}\;,
\end{equation}
where, as for single-layer graphene, $\Delta$ is neglected. Similarly to single-layer graphene, the conduction and valence band is symmetric, i.e., $E_c^{\stext{H}}=-E_v^{\stext{H}}$, resulting in inter-Landau-level transition energies \textit{independent} of the handedness of the circular polarization.

For the K-point series, the eigenvalues of the Hamiltonian for the conduction and valance band exhibit a similar behavior as for coupled graphene layers (Eqn.~\ref{eqn:LLNLayer}). In the \hyperref[appendix:SWC-model]{appendix} we derive an approximation up to the second order in $B_{\perp}$, where the linear part is given here
\begin{equation}
\begin{split}
\label{eqn:LL_graphite}
	E_{\stackrel{c}{\text{\tiny{\textit{v}}}}}^{\stext{K}}=
	&\pm\hbar\omega_{\stackrel{c}{\text{\tiny{\textit{v}}}}}\left(n+\frac{1}{2}\right)\\
	= & \pm \hbar\omega^*\left(n+\frac{1}{2}\right)\left(1\pm 4\eta\right)\;,
\end{split}
\end{equation}
with $\omega_{\stackrel{c}{\text{\tiny{\textit{v}}}}}=\frac{e B_{\perp}}{m_{\stackrel{c}{\text{\tiny{\textit{v}}}}}} = \frac{e B_{\perp}}{m^{*}}\left(1\pm 4\eta\right) = \omega^*\left(1\pm 4\eta\right)$, where $\omega_{\stackrel{c}{\text{\tiny{\textit{v}}}}}$, $\omega^*$, $m_{\stackrel{c}{\text{\tiny{\textit{v}}}}}$, $m^*$ and $\eta$ denote the cyclotron frequency of the conduction and valence band, the average cyclotron frequency, the effective mass of the conduction and valence band, the average effective mass and the first order splitting parameter, respectively, where $\eta$ and $m^*$ are determined by multiple band parameters\cite{PhysRevB.21.827} (see \hyperref[appendix:SWC-model]{appendix}). Note, the Landau level spectrum (Fig.~\hyperref[fig:Graphene_graphite_principle_LL]{\ref*{fig:Graphene_graphite_principle_LL} a}) of the conduction and valence band is asymmetric, i.e., $E_c^{\stext{K}}\neq-E_v^{\stext{K}}$, resulting in inter-Landau-level transition energies \textit{dependent} on the handedness of the circular polarization of the electromagnetic wave (photon) associated with the respective transition. For transitions with quantum numbers higher than $N\geq 5$, and therefore higher energies, non-parabolic band structure effects cause a deviation from the linear behavior. Further details are given in the \hyperref[appendix:SWC-model]{appendix}.

\subsection{Generalized Ellipsometry, optical Hall effect}\label{sec:OHE}
Generalized (Mueller matrix) ellipsometry is a polarization sensitive technique, extending spectroscopic (isotropic) ellipsometry,\cite{Azzam95,Fujiwara_2007} allowing to investigate arbitrary anisotropic and depolarizing materials. Mueller matrix ellipsometry can be used to determine a material's full ranked dielectric tensor. The Mueller matrix $\mathbf{M}$ is a real-valued $4\times 4$ transformation matrix for Stokes vectors~$\mathbf{S}$
\begin{equation}
	S^{(\text{out})}_j=\sum^{4}_{i=1}M_{ij}S^{(\text{in})}_i,\;\;(j=1\ldots 4)\;,
	\label{eqn:multi_MM}
\end{equation}
where $\mathbf{S}^{(\text{out})}$ and $\mathbf{S}^{(\text{in})}$ denote the Stokes vectors of the electromagnetic plane wave before and after the interaction with a sample, respectively.
\ifthenelse{\boolean{ext_bbl}}{\cite{Note_PRB4}}{\footnote{Note that all Mueller matrix elements of the GE data discussed in this paper, are normalized by the element $M_{11}$, therefore $|M_{ij}|\leq 1$ and therefore $M_{11}=1$.}} 
In terms of the \textit{p}- and \textit{s}-coordinate system
\ifthenelse{\boolean{ext_bbl}}{\cite{Note_PRB5}}{\footnote{\textit{p} and \textit{s} denote the directions parallel and perpendicular to the plane of incidence.}} 
the Stokes vector $\mathbf{S}$ can be defined as $S_1=I_p+I_s$, $S_2=I_p-I_s$, $S_3=I_{45}-I_{-45}$, and $S_4=I_{\sigma+}-I_{\mathit{\sigma}-}$, where $I_p$, $I_s$, $I_{45}$, $I_{-45}$, $I_{\sigma+}$, and $I_{\sigma-}$ are the intensities of linear \textit{p}- and \textit{s}-polarized, linear +45$^\circ$ and -45$^\circ$ polarized, and right- and left-handed circularly polarized light, respectively.\cite{AzzamBook_1984,Roseler90}

The Mueller matrix of a sample that is composed of a sequence of $k$ homogeneous layers with smooth and parallel interfaces can be calculated from the dielectric tensors $\eps^{(k)}$ of all $k$ layers using the $4\times 4$ matrix algorithm.~\cite{Schubert96,SchubertIRSEBook_2004,Fujiwara_2007} Employing the $4\times 4$ matrix algorithm it can be shown that if, and only if, at least one sample material's dielectric tensor possesses non-vanishing off-diagonal elements \textit{p}-\textit{s}-polarization mode conversion appears.
\ifthenelse{\boolean{ext_bbl}}{\cite{Note_PRB6}}{\footnote{\textit{p}-\textit{s}-polarization mode conversion appears if electromagnetic energy is converted from the parallel to the plane of incidence polarized channel (\textit{p}) to the perpendicular to the plane of incidence polarized channel (\textit{s}), or vice versa.}} 
Only then elements in the two off-diagonal-blocks
$
\text{
	\scriptsize
	$
		\begin{bmatrix}
			M_{13}&\hspace{-3pt}M_{14}\\
			M_{23}&\hspace{-3pt}M_{24}\\
		\end{bmatrix}
	$
	\normalsize
	and
	\scriptsize
	$
		\begin{bmatrix}
			M_{31}&\hspace{-3pt}M_{32}\\
			M_{41}&\hspace{-3pt}M_{42}\\
		\end{bmatrix}
	$
}
$
of the Mueller matrix of the sample deviate from zero. In the following we will use the terminology "\textit{polarization mode mixing}" when experimental data contains signatures from non-vanishing off-diagonal elements in the dielectric tensor and we will otherwise use the term "\textit{polarization mode conserving}".

Hereby we define \textbf{OHE data} $\mathbf{M}^\pm$ as Mueller matrix data from a generalized ellipsometry experiment with the sample exposed to an external, constant magnetic field $\pm \mathbf{B}$
\begin{equation}\label{OHE-MM-approx}
\begin{split}
	\mathbf{M}^\pm
	&=
		\mathbf{M}(\eps_{_{\hspace{-1pt}\pm\mathbf{B}}}^{(k)})\\
	&=
	\mathbf{M}(
	  \mathbf{I}
		+
		\bm{\chi}_{_{\hspace{-1pt}\mathbf{B}=0}}^{(k)}
		+\;
		\bm{\chi}_{_{\hspace{-0pt}\pm\mathbf{B}}}^{(k)})\;,
\end{split}
\end{equation}
where the Mueller matrix $\mathbf{M}^\pm$ is a function of all $k$, in general $\mathbf{B}$ dependent, dielectric tensors $\eps_{_{\hspace{-1pt}\pm\mathbf{B}}}^{(k)}$ of a sample with $k$ optically distinguishable
\ifthenelse{\boolean{ext_bbl}}{\cite{Note_PRB7}}{\footnote{Note, that the individual graphene layers of $m$-layer graphene are optically indistinguishable in this sense, since they can be described by the same dielectric tensor.}} 
layers.

Additionally, we define \textbf{OHE difference data} $\delta\mathbf{M}^{\pm}$ as the difference data between two Mueller matrix datasets, measured at the magnetic field $\pm \mathbf{B}$ and the corresponding zero field dataset
\begin{equation}\label{delta-MM-pm}
	\begin{split}
		\delta\mathbf{M}^{\pm}
		&=
		\mathbf{M}^\pm
		\hspace{-2pt}-\hspace{-2pt}
		\mathbf{M}_0\\
		&=
	\mathbf{M}(
	  \mathbf{I}
		+
		\bm{\chi}_{_{\hspace{-1pt}\mathbf{B}=0}}^{(k)}
		+
		\bm{\chi}_{_{\hspace{-0pt}\pm\mathbf{B}}}^{(k)})
		-
			\mathbf{M}(
	  \mathbf{I}
		+
		\bm{\chi}_{_{\hspace{-1pt}\mathbf{B}=0}}^{(k)})\;.
	\end{split}
\end{equation}
where $\mathbf{M}_0$ is the Mueller matrix measured at $\mathbf{B}=0$.

Mueller matrix (optical Hall effect) datasets depend on extrinsic parameters, such as angle of incidence, temperature or magnetic field, as well as intrinsic parameters, such as sample structure, carrier concentration or mobility. To extract physical relevant information from Mueller matrix datasets, optical models and nonlinear regression methods are used to analysis data. A detailed description of the OHE data analysis is omitted here for brevity and the interested reader is referred to Ref.~\citenum{KuehneRSI2014} and references therein.

\begin{figure*}[htbp]
	\centering
  \ifthenelse{\boolean{eV}}{
	\includegraphics[
	width=0.98\textwidth]{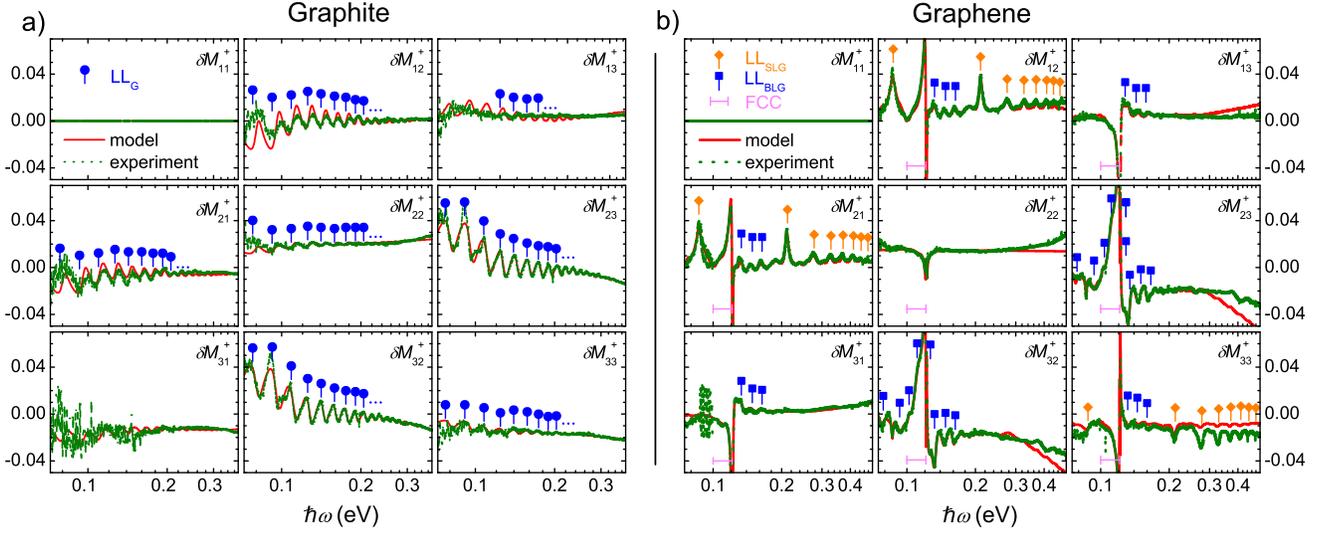}}{
	\includegraphics[
	width=0.98\textwidth]{graphite_graphene-8T.eps}}	
	\caption{Experimental (green, dashed lines) and best-match model calculated (red, solid lines) using the small energy splitting approximation (Eqn.~\ref{eqn:landau_averaged}). OHE difference datasets $\delta\mathbf{M}^{+}$ at $T=1.5$~K and $B=8$~T ($B_{\perp}=5.66$~T) for a) graphite and b) epitaxial graphene at $\Phi_{\text{a}}=45^{\circ}$ angle of incidence. For graphite K-point inter-Landau-level transitions (LL$_{\text{G}}$) are found to be polarization mode mixing, while inter-Landau-level transition in the epitaxial graphene sample can be separated in polarization mode preserving (LL$_{\text{SLG}}$) and polarization mode mixing (LL$_{\text{BLG}}$).}
	\label{fig:graphite+,-}
\end{figure*}

\begin{figure*}[htbp]
	\centering
  \ifthenelse{\boolean{eV}}{
  \includegraphics[
	width=0.98\textwidth]{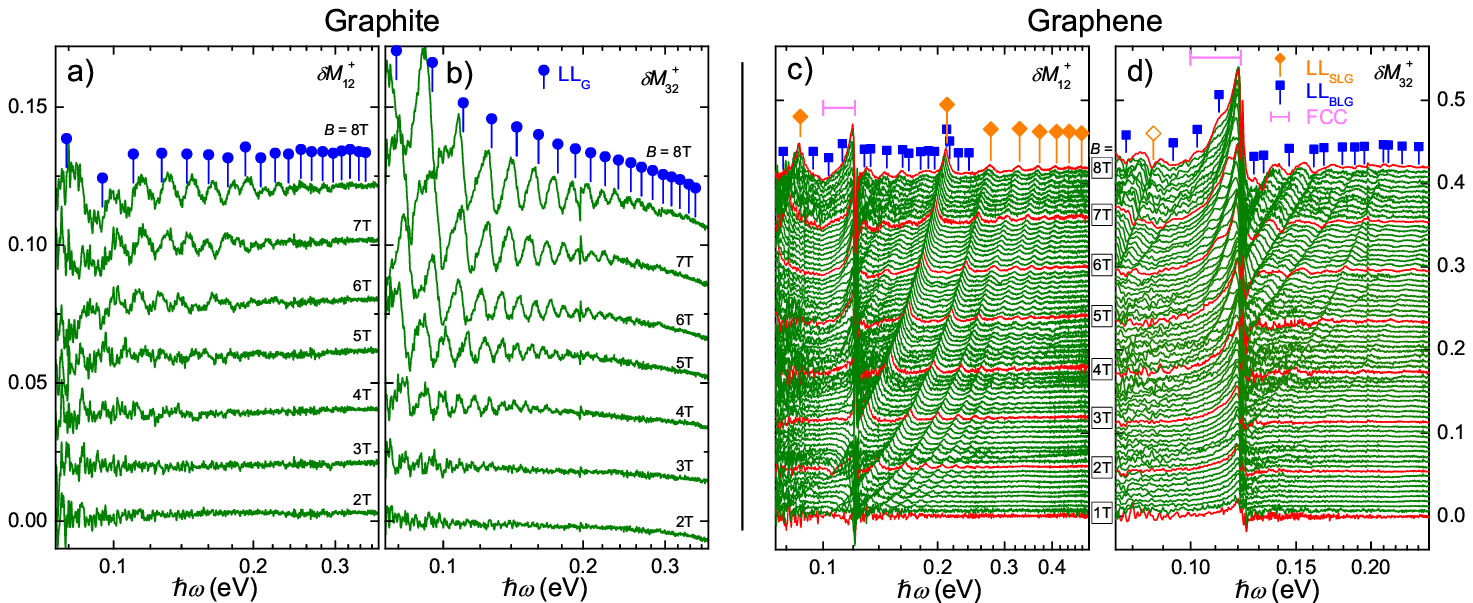}}{
	\includegraphics[
	width=0.98\textwidth]{graphite_graphene-field_dependence.eps}}
	\caption{Magnetic field dependency of selected, representative block-on- a, c) and block-off-diagonal elements b, d) $\delta M^{+}_{12}$ and $\delta M^{+}_{32}$, respectively, of the experimental OHE difference datasets at $T=1.5$~K and $\Phi_{\text{a}}=45^{\circ}$, for a, b) graphite and c, d) epitaxial graphene.
	Inter-Landau-level transitions with fundamentally different $B_{\perp}$-dependency and polarization selection rules are observed in graphite (labeled LL$_\text{G}$: polarization mode mixing) and epitaxial graphene (labeled LL$_{\text{SLG}}$: polarization mode conserving; labeled LL$_{\text{BLG}}$: polarization mode mixing). For epitaxial graphene, inter-Landau-level transitions labeled LL$_{\text{SLG}}$ can be assigned to single-layer graphene or multi layer graphene with an odd layer number $m$ and transitions labeled LL$_{\text{BLG}}$ can be assigned to multilayer graphene. The magneto-optical response of free charge carriers can be observed as a slope in $\delta M^{+}_{32}$ for graphite and a long-wavelength interface mode for graphene (labeled FCC). For graphite the magnetic field ranges from $B=2$~T to 8~T in 1~T increments, and for epitaxial graphene from $B=1$~T to 8~T in 0.1~T increments (note: $B_{\perp}=B/\sqrt{2}$). The graphs are stacked by 0.02 for graphite and 0.006 for graphene and red lines were used as a guide to the eye for $B=1, 2\dots 8$T in case of epitaxial graphene.}
	\label{fig:graphite-graphene_field_dependence}
\end{figure*}

\section{Experiment}
Two samples were investigated, a highly oriented pyrolytic graphite~(HOPG) sample and a multilayer graphene on SiC sample. The graphite sample (Advanced Ceramics) has a rocking curve peak with FWHM~$\approx 0.4^{\circ}$.\cite{Kempa02} 
The multilayer graphene was grown on the C-polar (000$\bar{1}$) surface of a semi-insulating 6\textit{H}-SiC substrate. During the growth of the epitaxial graphene, the SiC substrate was heated to 1400~$^\circ$C in an argon atmosphere, resulting in sublimation of silicon atoms from the surface. Further information on growth conditions can be found in Ref.~\onlinecite{TedescoAPL96_2010}. The number of graphene layers is estimated to be 10-20, similar to those measured previously on C-face 4\textit{H}-SiC~\cite{BoosalisAPL101_2012}. Mobility and sheet charge carrier density of the epitaxial graphene at room temperature were determined by electrical Hall effect measurements as $\mu=1714$~cm$^2$/Vs and $N_s=6.7\times 10^{13}$~cm$^{-1}$, respectively.

For both samples optical Hall effect measurements where carried using the mid-infrared sub-system of the  integrated mid-infrared, far-infrared, and terahertz optical Hall effect instrument.\cite{KuehneRSI2014} All measurements were carried out at a temperature of $T=1.5$~K in the spectral range from 74 to 500~meV ($600-4000$~cm$^{-1}$) with a spectral resolution of 0.12~meV (1~cm$^{-1}$).\cite{KuehneRSI2014,HofmannRSI77_2006} The magnetic field was varied from $B=0$~T to 8~T in 1~T increments for the graphite sample and in 0.1~T increments for the epitaxial graphene sample. For graphite data was also recorded for both magnetic field directions. The angle of incidence during the optical Hall effect measurements was $\mathit{\Phi}_a$~=~45$^\circ$, while the magnetic field direction was parallel to the reflected beam. According to Refs.~\onlinecite{doi:10.1143/JPSJ.33.1619} and \onlinecite{PhysRev.134.A453}, the effect from the direction of the magnetic field $B$ on the inter-Landau-level transitions only shows a $1/\cos{\theta}$-dependence, where $\theta$ is the angle between the magnetic field and the c-axis of the graphite crystal. Therefore we use for all calculations the magnetic field component $B_{\perp}=B$/$\sqrt{2}$ along the sample normal and parallel to the c-axis of graphite and multilayer graphene.

\section{Results and Discussion}
Figure~\ref{fig:graphite+,-} depicts experimental OHE difference data $\delta\mathbf{M}^{+}$ (green dotted lines), recorded at $\Phi_{\text{a}}=45^{\circ}$ angle of incidence, for $B=+8$~T ($B_{\perp}=+5.66$~T) and $T=1.5$~K, and best model calculations (red solid lines) for graphite (Fig.~\hyperref[fig:graphite+,-]{\ref*{fig:graphite+,-}a}) and multilayer graphene (Fig.~\hyperref[fig:graphite+,-]{\ref*{fig:graphite+,-}b}). Contributions from free charge carriers as well as from inter-Landau-level transitions to the magneto-optical response are detected. For both samples the strongest magnetic field induced change in the Mueller matrix $\delta\mathbf{M}^{\pm}$ is due to free charge carriers, and can, in case of graphite, be observed as a slope in the matrix elements $\delta M^{+}_{23}$ and $\delta M^{+}_{32}$, while for multilayer graphene a strong feature between the TO and LO phonon frequency of the SiC substrate is detected (Fig.~\hyperref[fig:graphite+,-]{\ref*{fig:graphite+,-}b}, \hyperref[fig:graphite-graphene_field_dependence]{\ref*{fig:graphite-graphene_field_dependence}c} and \hyperref[fig:graphite-graphene_field_dependence]{d}, labeled as FCC). For graphite, signatures from inter-Landau-level transitions (Fig.~\hyperref[fig:graphite+,-]{\ref*{fig:graphite+,-}a}, \hyperref[fig:graphite-graphene_field_dependence]{\ref*{fig:graphite-graphene_field_dependence}a} and \hyperref[fig:graphite-graphene_field_dependence]{\ref*{fig:graphite-graphene_field_dependence}b}, labeled as LL$_{\text{G}}$) are observed in all Mueller matrix elements indicating a polarization mixing behavior. For multilayer graphene signatures from a set of inter-Landau-level transitions (Fig.~\hyperref[fig:graphite+,-]{\ref*{fig:graphite+,-}b}, \hyperref[fig:graphite-graphene_field_dependence]{\ref*{fig:graphite-graphene_field_dependence}c} and \hyperref[fig:graphite-graphene_field_dependence]{d}, labeled as LL$_{\text{BLG}}$) appear in all Mueller matrix elements and therefore possesses polarization mixing characteristics, while signatures from a second set of transitions (labeled as LL$_{\text{SLG}}$) is only observed in the on-diagonal-blocks,
$
\text{
	\scriptsize
	$
		\begin{bmatrix}
			\delta M_{11}^{+}&\hspace{-3pt}\delta M_{12}^{+}\\
			\delta M_{21}^{+}&\hspace{-3pt}\delta M_{22}^{+}\\
		\end{bmatrix}
	$
	\normalsize
	and
	\scriptsize
	$
		\begin{bmatrix}
			\delta M_{33}^{+}&\hspace{-3pt}\delta M_{34}^{+}\\
			\delta M_{43}^{+}&\hspace{-3pt}\delta M_{44}^{+}\\
		\end{bmatrix}
	$
}
$, 
of the OHE difference data.
\ifthenelse{\boolean{ext_bbl}}{\cite{Note_PRB8}}{\footnote{Note that the ellipsometer used here is a rotating analyzer ellipsometer and cannot measure the last row and column of the Mueller matrix.}} 
The latter set therefore possesses polarization preserving characteristics.\cite{KuehnePRL111_2013}


\subsection{Magnetic field dependence}
In order to determine the physical origin of the different types of inter-Landau-level transitions, magnetic field dependent measurements were conducted (Fig.~\ref{fig:graphite-graphene_field_dependence}). Figures~\hyperref[fig:graphite-graphene_field_dependence]{\ref*{fig:graphite-graphene_field_dependence}a} and \hyperref[fig:graphite-graphene_field_dependence]{\ref*{fig:graphite-graphene_field_dependence}c} show the magnetic field dependencies of $\delta M_{12}^{+}$, a representative block on-diagonal element, while figures~\hyperref[fig:graphite-graphene_field_dependence]{\ref*{fig:graphite-graphene_field_dependence}b} and \hyperref[fig:graphite-graphene_field_dependence]{\ref*{fig:graphite-graphene_field_dependence}d} show the magnetic field dependencies of $\delta M_{32}^{+}$, a representative block off-diagonal element, for graphite (figures~\hyperref[fig:graphite-graphene_field_dependence]{\ref*{fig:graphite-graphene_field_dependence}a,b}) and multilayer graphene (figures~\hyperref[fig:graphite-graphene_field_dependence]{\ref*{fig:graphite-graphene_field_dependence}c,d}). For both samples the magnetic field dependencies of the free charge carrier effects (best observed in $\delta M^{+}_{32}$), vary with the external magnetic field only in magnitude but not in spectral position, resulting in a change of the underlying slope in Fig.~\hyperref[fig:graphite-graphene_field_dependence]{\ref*{fig:graphite-graphene_field_dependence}b} and a change in amplitude for the FCC labeled feature in Fig.~\hyperref[fig:graphite-graphene_field_dependence]{\ref*{fig:graphite-graphene_field_dependence}d}. This can be explained using Eq.~\ref{eqn:chi_drude_pm} under consideration that, for the applied external magnetic fields, the cyclotron frequency is well below the experimental spectral range ($\omega>>\omega_c$). As a result the on-diagonal elements in Eq.~\ref{eqn:general_susceptibility} can be approximated by $\chi_{jj}=\chi^{\text{\tiny{D}}}$ (Eq.~\ref{eqn:iso_drude}) and the off-diagonal elements are proportional to the cyclotron frequency $\chi_{ij}\propto\pm\text{i}\chi^{\text{\tiny{D}}}\frac{\omega_c}{\omega+\text{i}\gamma}$, and therefore scale linearly with the magnetic field. Further, the free charge carrier contributions are observed as a strong feature in the spectral range where the phonon modes of the SiC substrate occur, which can be attributed to a long-wavelength interface mode between the SiC substrate and the graphene layer.\cite{SchubertPRB71_2005}

\begin{figure*}[htbp]
	\centering
  \ifthenelse{\boolean{eV}}{
  \includegraphics[
	width=0.98\textwidth]{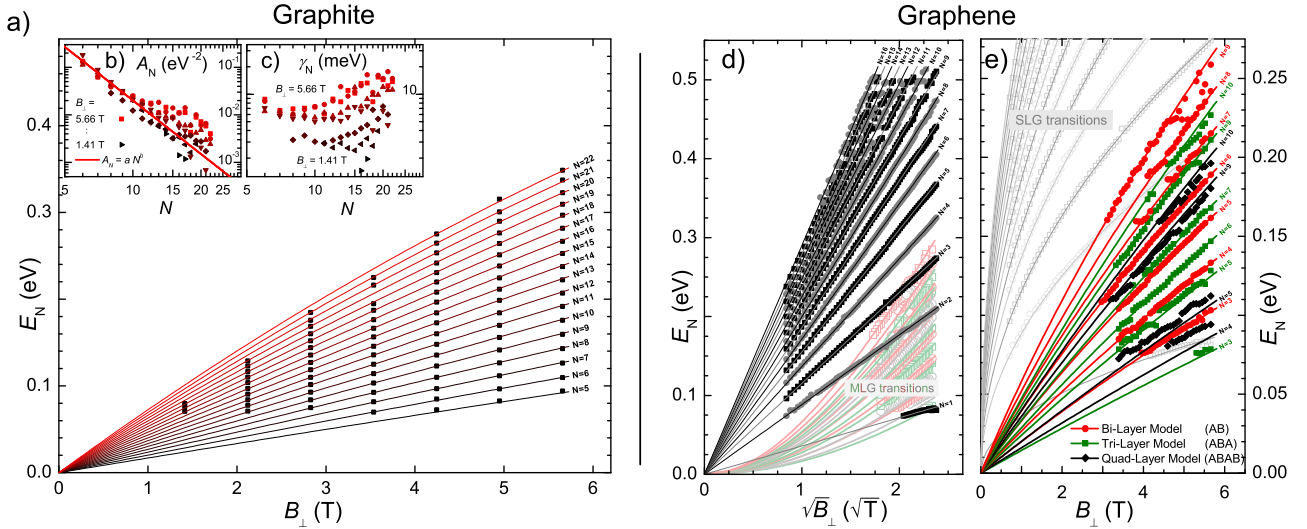}}{
  \includegraphics[
	width=0.98\textwidth]{Transition_energies3.eps}}
	\caption{Symbols: Best model inter-Landau-level transition parameters from analysis of OHE data using the small energy splitting approximation, where figures a, d, e) display transition energies, b) amplitudes and c) broadening parameters for graphite a, b, c) and multilayer graphene d, e).
	Figures d) and e) depict the magnetic field dependencies of best-match model energy parameters for polarization mode conserving and polarization mode mixing inter-Landau-level transitions in epitaxial graphene, respectively,
	where d) is plotted vs. $\sqrt{B_{\perp}}$ and e) vs. $B_{\perp}$. The $\sqrt{B_{\perp}}$-dependency in d) indicates that these inter-Landau-level transitions either occur in single-layer graphene or in coupled, Bernal stacked graphene layers with an odd layer number $m$.\cite{KoshinoPRB77_2008}
	The best-match model parameter for the Fermi velocity is determined as $v_f=(1.01\pm0.01)\times 10^6$m/s.
	In figure e) energy parameters plotted as red circles, green squares and black diamonds are assigned to Bernal stacked bi-, tri- and quad-layer graphene, respectively. Solid lines represent best model calculations using the Fermi velocity determined from polarization mode conserving inter-Landau-level transitions. The best-match model parameter for the inter-layer coupling constant for bi-, tri- and quad-layer graphene are
$\gamma_1^{(2)}=(0.387\pm 0.001)$eV, $\gamma_1^{(3)}=(0.385\pm 0.001)$eV and $\gamma_1^{(4)}=(0.405\pm 0.003)$eV, respectively.
}
	\label{fig:Graphene_transition_energies_0}
\end{figure*}

For graphite, the spectral positions of the polarization mode mixing inter-Landau-level transitions LL$_{\text{G}}$ are found to scale sub-linearly with $B_{\perp}$ and can be assigned to the K-point inter-Landau-level transitions in graphite.
\ifthenelse{\boolean{ext_bbl}}{\cite{Note_PRB9}}{\footnote{Sub-linear behavior means $B_{\perp}^x$ with $1> x > \frac{1}{2}$, and can be observed comparing the line shapes in Fig.~\hyperref[fig:Graphene_transition_energies_0]{\ref*{fig:Graphene_transition_energies_0}d}~and~\hyperref[fig:Graphene_transition_energies_0]{e}}} 
H-point transitions are not observed at first glance, however, as discussed below, can be identified in a more detailed analysis.

For multilayer graphene on SiC, the polarization mode mixing set of inter-Landau-level transitions LL$_{\text{BLG}}$  exhibits a sub-linear $B_{\perp}$-dependency similar to graphite. Due to their magnetic field dependence, these transitions can be assigned to the bi-layer branch of $m$-layer graphene.\cite{KoshinoPRB77_2008} The other set of inter-Landau-level transitions LL$_{\text{SLG}}$ exhibits a $\sqrt{B_{\perp}}$-dependency and is therefore assigned to single-layer graphene and multilayer graphene with an odd number of layers.\cite{KoshinoPRB77_2008} The inter-Landau-level transitions LL$_{\text{SLG}}$ are polarization mode conserving, with exception of the lowest observed transition with $N=1$ ($\hbar\omega \approx$ 86~meV). The $N=1$ transition is polarization mode mixing in its appearance within the OHE data (see Fig.~\hyperref[fig:graphite-graphene_field_dependence]{\ref*{fig:graphite-graphene_field_dependence}d}, hollow orange diamond).

\subsection{Polarization selection rules}
Polarization state conserving and mixing behavior for inter-Landau-level transitions originates from symmetric, i.e., diagonal, and antisymmetric, i.e., non-diagonal magnetic field induced electric susceptibility tensors $\bm{\chi}_{_{\hspace{-1pt}\mathbf{B}}}$ (Eqn.~\ref{eqn:general_susceptibility}), respectively. The properties of $\bm{\chi}_{_{\hspace{-1pt}\mathbf{B}}}$ are determined by the band symmetry or asymmetry of the valence and conduction band. In general individual transitions between Landau levels involve absorption of circularly polarized light.
\ifthenelse{\boolean{ext_bbl}}{\cite{Note_PRB10}}{\footnote{This includes higher order optical selection rules with $n'\neq n\pm 1$\cite{PhysRev.140.A401}, which are neglected here.}} 
Therefore, individual inter-Landau-level transitions $\text{L}_n^-\rightarrow \text{L}_{n'}^+$ without their counterpart $\text{L}_{n'}^-\rightarrow \text{L}_{n}^+$, always result in a antisymmetric electric susceptibility tensor $\bm{\chi}_{_{\hspace{-1pt}\mathbf{B}}}$.

For graphite, at the K-point, the Landau level spectrum of the conduction and valence band is asymmetric, i.e., $E_c^{\stext{K}}\neq-E_v^{\stext{K}}$, leading to a different spacing of the Landau levels in the valence and conduction band. Therefore, the transitions $\text{L}_n^-\rightarrow \text{L}_{n+1}^+$ and $\text{L}_{n+1}^-\rightarrow \text{L}_n^+$ are not equivalent in terms of their transition energies. Even if the broadening and amplitudes are equal, the susceptibilities for right- and left-handed circular polarized light are different, \mbox{$\chi_{+}\neq \chi_{-}$}, and therefore, $\bm{\chi}_{_{\hspace{-1pt}\mathbf{B}}}$ becomes antisymmetric (non-diagonal). At the H-point, the Landau level spectrum is symmetric, i.e., $E_c^{\stext{H}} = -E_v^{\stext{H}}$, resulting in the transitions $\text{L}_n^-\rightarrow \text{L}_{n+1}^+$ and $\text{L}_{n+1}^-\rightarrow \text{L}_n^+$ having the same transition energy. Using Eqn.~\ref{eqn:landau}, under the assumption that broadening and amplitude are equal as well, the susceptibilities for right- and left-handed circular polarized light are equal, \mbox{$\chi_{+}=\chi_{-}$}, and therefore, $\bm{\chi}_{_{\hspace{-1pt}\mathbf{B}}}$ is symmetric (diagonal).\cite{PhysRevB.84.235410}

In case of single-layer graphene, the Landau level spectrum is symmetric in the conduction and valence band, equivalent to the H-point Landau level spectrum of graphite, i.e., $E_c^{\stext{SLG}}=-E_v^{\stext{SLG}}$, resulting in the transitions $\text{L}_n^-\rightarrow \text{L}_{n+1}^+$ and $\text{L}_{n+1}^-\rightarrow \text{L}_n^+$ having the same transition energy. Therefore, under the assumption of equal broadening and amplitude, $\bm{\chi}_{_{\hspace{-1pt}\mathbf{B}}}$ for single-layer graphene is symmetric (diagonal).

The polarization mixing behavior of the lowest observed SLG inter-Landau-level transition with $N=1$ can be explained by the position of the Fermi level $E_{\text{F}}$.\cite{PhysRevB.84.235410} Here, $E_{\text{F}}$ lies between the $\text{L}_0^\pm$ and the $\text{L}_1^\pm$ Landau level but is not within a few $k_B T$ to any of these levels. This leads to fully occupied or fully empty initial and end Landau levels for one of the $N=1$ inter-Landau-level-transitions. Thus, one of the two $N=1$ transition is not permitted (see Fig.~\hyperref[fig:Graphene_graphite_principle_LL]{\ref*{fig:Graphene_graphite_principle_LL}b}), which results in absorption of right- or left-handed polarized light only, i.e., $\chi_{+}\neq \chi_{-}$, and therefore to an antisymmetric (non-diagonal) contribution to the electric susceptibility tensor. When shifting the Fermi-level across the $\text{L}_0^\pm$ Landau level, which is equivalent to a change from p- to n-type doping or vice versa, $\chi_{+}$ and $\chi_{-}$ interchange and the feature for the $N=1$ transition changes sign in the off-diagonal-block elements of the OHE data. From the experimental OHE data we determine that our epitaxial graphene sample is n-type doped.

The experimental OHE data shows that inter-Landau-level transitions, originating from the bi-layer branch of $m$-layer graphene, possess polarization mixing behavior (BLG). Therefore, we can conclude that in this case \mbox{$\chi_{+}\neq \chi_{-}$}. According to Eqn.~\ref{eqn:landau}, in order to account for different electric susceptibilities for right- and left-handed circular polarized light, at least one of the three parameters amplitude, broadening and transition energy must be different for the two circular polarized absorptions. Similar to the case of graphite, we attribute the polarization mixing behavior to an energy splitting between the two inter-Landau-level transitions with opposing circular polarization, which is not included in Eqn.~\ref{eqn:LLNLayer} due to the simplifications \mbox{$\gamma_3=0$} and \mbox{$\gamma_4=0$}. Equation~\ref{eqn:LLNLayer} therefore calculates the central inter-Landau-level transitions energies of the small energy splitting approximation.

\begin{table*}[tb]
	\centering
	\caption{best-match model calculated tight binding parameters for the SWM model for graphite and multilayer graphene. For graphene, the upper index denotes the number of coupled graphene layers. All values are given in $[$eV$]$.}
		\begin{tabular}{l c c c c c }
		\hline
	\noalign{\vskip 2pt}
			\multirow{3}{*}{\parbox[c]{2cm}{Graphite}} & $\gamma_0$ & $\gamma_1$ & $\gamma_3$ & $\gamma_4$ & $\gamma_5-\gamma_2-\Delta/2$ \\
	\noalign{\vskip 2pt}
		\cline{2-6}
	\noalign{\vskip 2pt}
			 & \parbox[c]{2.5cm}{$3.12$
\footnote{From Ref.~\onlinecite{Chung2002}} (fixed\footnote{see Results and Discussion Sec.~\ref{sec:res_graphite}})} & \parbox[c]{2.5cm}{$0.395\pm 0.006$} & \parbox[c]{2.5cm}{0 (fixed)} & \parbox[c]{2.5cm}{$0.10\pm 0.02$} & \parbox[c]{2.5cm}{$0.063\pm 0.013$}\\
	\noalign{\vskip 2pt}
\hline
\noalign{\vskip 5pt}
\cline{1-6}
\noalign{\vskip 2pt}
		\multirow{2}{*}{\parbox[c]{2cm}{Graphene}} & $\gamma_0^{(1)}$ & $\gamma_1^{(2)}$ & $\gamma_1^{(3)}$ & $\gamma_1^{(4)}$ & $\gamma_4^{(2)}$ \\
\noalign{\vskip 2pt}
		\cline{2-6}
	\noalign{\vskip 2pt}
		& \parbox[c]{2.5cm}{$3.12\pm 0.03$} & \parbox[c]{2.5cm}{$0.387\pm 0.001$} & \parbox[c]{2.5cm}{$0.385\pm 0.001$} & \parbox[c]{2.5cm}{$0.405\pm 0.003$} &  \parbox[c]{2.5cm}{$0.23\pm 0.06$}\\
	\noalign{\vskip 2pt}
		\cline{1-6}
		\end{tabular}
	\label{tab:BestModelSWMGraphite}
\end{table*}

\subsection{Optical model and data analysis}
For graphite, the optical model consists of a single, semi-infinite layer, whose dielectric tensor contains contributions from inter-Landau-level transitions and free charge carriers according to Eqn.~\ref{eqn:fulldrude}. 

For multilayer graphene, the optical model consists of two layers, a semi-infinite layer for the 6$H$-SiC substrate and a layer with fixed thickness of $d=1$~nm for the graphene. The substrate is modeled using a uniaxial dielectric tensor according to Eqn.~\ref{eqn:phonon2} for a single TO-LO phonon resonance ($l=1$). The graphene layer contains contributions from inter-Landau-level transitions and free charge carriers according to Eqn.~\ref{eqn:fulldrude}. 

For both samples, Eqn.~\ref{eqn:landau} with $A_{+,N}=A_{-,N}=A_{N}$, $E_{+,N}=E_{-,N}=E_{N}$ and $\gamma_{+,N}=\gamma_{-,N}=\gamma_{N}$ (for $N\geq 2$), resulting in a symmetric (diagonal) electric susceptibility tensor $\bm{\chi}_{_{\hspace{-1pt}\mathbf{B}}}$, is used to describe the polarization mode conserving inter-Landau-level transitions.

For polarization mode mixing inter-Landau-level transitions, the relative strength of signatures in the on- and off-diagonal blocks of $\delta\mathbf{M}^{\pm}$ depends on the ratio of the on- and off-diagonal elements of $\bm{\chi}_{_{\hspace{-1pt}\mathbf{B}}}$. This allows to distinguish between $\chi_{+}$ and $\chi_{-}$.
\ifthenelse{\boolean{ext_bbl}}{\cite{Note_PRB11}}{\footnote{Note that the experimental setup does contain linear polarizers only.}} 
For the data presented here, the limitations in distinguishing between $\chi_{+}$ and $\chi_{-}$ during the model based analysis, stem from the signal-to-noise ratios for the signatures from inter-Landau-level transitions in the on- and off-diagonal blocks of $\delta\mathbf{M}^{\pm}$. For both samples, for the highest field measured, the signal-to-noise ratio in the off-diagonal block elements (i.e., $\delta M_{23}^{\pm}$ and $\delta M_{32}^{\pm}$) is approximately 15, while the on-diagonal block elements (i.e., $\delta M_{12}^{\pm}$ and $\delta M_{21}^{\pm}$) exhibit a signal-to-noise ratio of approximately 3. These ratios reduce with the decreasing magnetic field strength, resulting, for the OHE datasets measured here, in a loss of sensitivity to model parameters for magnetic fields below approx. $B=7$~T. Therefore, we used in this paper, for both sample systems, the \textit{small energy splitting approximation} (Eqn.~\ref{eqn:landau_averaged}) for data analysis of polarization mode mixing sets of inter-Landau-level transitions. Only for the highest magnetic field measured, we applied the \textit{circular polarization transition model} (Eqn.~\ref{eqn:landau}).

\subsection{Band structure parameters}

\subsubsection{Graphite}
The free charge carrier contribution to the magneto-optical response of graphite was modeled isotropic and the model parameters were the same for all magnetic fields. The effective mass was coupled to the average effective mass of free charge carriers $m^*=0.068\: m_e$ (see below) and the best-match-model parameters for the free charge carrier concentration and mobility parameters are $n= (5.1 \pm 0.2)\times 10^{19}$cm$^{-3}$ and $\mu=(30\pm 3)$cm$^2$/Vs.

Symbols in Figs.~\hyperref[fig:Graphene_transition_energies_0]{\ref*{fig:Graphene_transition_energies_0}a, b} and \hyperref[fig:Graphene_transition_energies_0]{c} display best-match-model energy, amplitude, and broadening parameters, respectively, of the $N^{\text{th}}$ inter-Landau-level transition at the K-point, obtained from OHE data analysis using the small energy splitting model (Eqn.~\ref{eqn:landau_averaged}). Lines in Fig.~\hyperref[fig:Graphene_transition_energies_0]{\ref*{fig:Graphene_transition_energies_0}a} display the transition energies of the small energy splitting approximation $E^{\stext{K}}_{N}$ vs. $B_{\perp}$ (Eqn.~\ref{eqn:circ_av_model}), where the superscript K indicates here and in the following the K-point, i.e., $\Gamma=2$. First, the best-match-model parameter set of quantum numbers for the inter-Landau-level transitions was determined as $N=5\dots 22$. From the linear dependence on $B_{\perp}$ we evaluate the average effective mass of free charge carriers as \mbox{$m^*=(0.068\pm 0.001)m_{\text{e}}$}, which is in excellent agreement with the literature, e.g., Refs.~\onlinecite{PhysRevB.21.827,Zhou2006}. From the $B_{\perp}^2$ dependence of $E^{\stext{K}}_{N}$ we obtain $\frac{3\eta-\frac{\gamma_4}{\gamma_0}}{\gamma_1}=(0.420\pm 0.005)$eV$^{-1}$. It is worth noting that we excluded transitions with $N<5$ in the analysis due to effects from the neglected band structure parameter $\gamma_3$.\cite{PhysRevB.21.827,doi:10.1143/JPSJ.40.761,Chung2002} The onset of the deviation of the inter-Landau-level-transition energies from Eqn.~\ref{eqn:landau_averaged} can be observed in Fig.~\hyperref[fig:Graphene_transition_energies_0]{\ref*{fig:Graphene_transition_energies_0}a} for the lowest transition $N=5$. Inserts \hyperref[fig:Graphene_transition_energies_0]{\ref*{fig:Graphene_transition_energies_0}b} and \hyperref[fig:Graphene_transition_energies_0]{c} show double-logarithmic plots of the best-match-model parameters from the OHE data analysis for amplitude and broadening, using the small energy splitting approximation. The amplitude parameter reveals for all fields an approximate an $A_{N}=a N^b$ dependency, while the broadening parameter increases with the magnetic field and also slightly with the quantum number.


For the highest magnetic field $B_{\perp}=8/\sqrt{2}$~T, we apply the circular polarization transition model (Eqn.~\ref{eqn:landau}). The transition energies for right- and left-handed circular polarized light are \mbox{$E^{\stext{K}}_{\pm,N}=E^{\stext{K}}_{N}\pm\frac{\delta E^{\stext{K}}_{N}}{2}$}, where we fixed $E^{\stext{K}}_{N}$ using the parameter determined by the small energy splitting approximation. The amplitude and broadening parameters for right- and left-handed circular polarized light were set equal $A_{\pm,N}=A_{N}$, and $\gamma_{\pm,N}=\gamma_{N}$, respectively. Further, the amplitude parameters were coupled to the transition number by $A_{N}=a N^b$, and the broadening parameters were set constant for all transitions ($\gamma_N=\gamma$). The best-match-model parameters are $a=0.30$~eV$^{-2}$, $b=-2.44$ and $\gamma=8.9$~meV. The model parameter $\delta E^{\stext{K}}_{N}$ for the energy splitting was linearized in the optical model $\delta E^{\stext{K}}_{N}=\delta E_0 + N\;\delta E_1$. The constant offset was determined as $\delta E_0=8\eta\hspace{3pt}\hbar\frac{e B}{m^*}=(5.06\pm 0.40)$meV and therefore $\eta=0.066 \pm 0.006$ (Eqn.~\ref{eqn:energy_splitting}), which is similar to values reported by Dresselhaus \textit{et.al.}\cite{PhysRevB.21.827}. Due to its large uncertainty, the best-match calculation has no sensitivity to the slope $\delta E_1=(-0.02\pm 0.06)$meV, and therefore, the band structure parameter $\gamma_1$ is not determined using Eqn.~\ref{eqn:energy_splitting}.

\begin{figure}[t]
	\centering
  \includegraphics[
	width=0.48\textwidth]{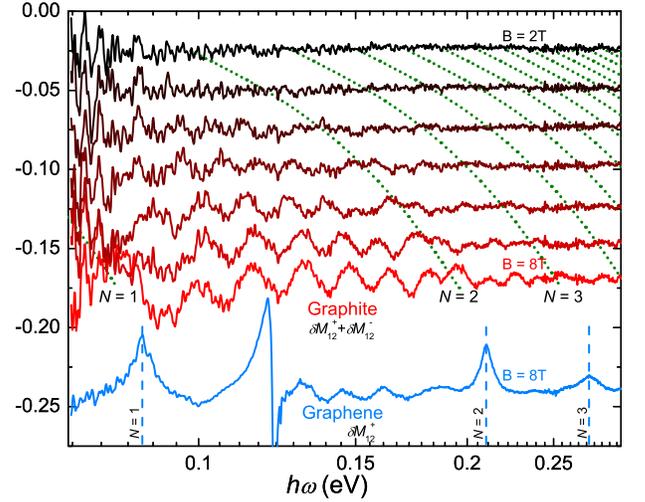}
	\caption{Upper lines: Magnetic field evolution ($B=8\dots 2$~T) of the on-diagonal block element $\delta M_{12}^{+}+\delta M_{12}^{-}$, displaying K-point inter-Landau-level transitions in graphite. Green dotted lines indicate the calculated magnetic field evolution with Fermi velocity $v_f=0.94\times 10^6$m/s ($\gamma_0=2.9$~eV) for H-point transitions.	Lower line: For comparison, $\delta M_{12}^{+}$ for $B=8$~T for graphene sample, displaying single-layer graphene inter-Landau-level transitions $N=1,2,3$. While inter-Landau-level transitions in single-layer graphene exhibit a clear resonance feature, H-point transitions in graphite are observed as a weak (with respect to the signal-to-noise ratio) and broad features.
	}
	\label{fig:Graphite_dmp+dmm}
\end{figure}

Direct access to the intra-layer coupling parameter $\gamma_0$ is provided by the H-point transitions (Eqn.~\ref{eqn:LLNLayer}). The problem arises here, in terms of sensitivity to $\gamma_0$, from the fact that the H-point transitions are polarization mode conserving (with exception of the $N=1$ transition) and therefore are only observed in the on-diagonal block Mueller matrix elements, which, as discussed above, have a low signal-to-noise ratio. Further, compared to the K-point transitions, the H-point transitions reveal a larger broadening and a more complex line shape, as discussed in detail in Ref.~\onlinecite{Toy77}. Attempts to determine $\gamma_0$ from the $N=1$ transition failed for multiple reasons. Even for the highest magnetic field measured, the $N=1$ transition is located at the lower edge of the spectral range measured (Fig.~\ref{fig:Graphite_dmp+dmm}), and therefore has an unfavorable signal-to-noise ratio. Due to its polarization mode mixing behavior, the $N=1$ H-point transition is indistinguishable from the K-point transitions, which, in addition, begin to be affected by the trigonal band warping ($\gamma_3$) and are therefore not be as well ordered as transition with a higher quantum number. A line shape analysis of $\delta\mathbf{M}^{+}+\delta\mathbf{M}^{-}$ (Fig.~\ref{fig:Graphite_dmp+dmm}) 
yielded $\gamma_0=(2.9\pm 0.1)$eV, which corresponds to a Fermi velocity of $v_f=(0.94\pm 0.01)\times 10^6$~m/s. However, applying the model for $m$-layer graphene $E_{N}^{\stext{m-BLG}}$ (Eqn.~\ref{eqn:LLNLayer}) for an infinite number of layers (\mbox{$m\rightarrow \infty$}) and $\Gamma=2$ for the analysis of the K-point transitions, a value of $\gamma_0=(3.01\pm 0.03)$eV, which corresponds to a Fermi velocity of $v_f=(0.98\pm 0.01)\times 10^6$~m/s, is obtained.\phantomsection\label{sec:res_graphite}

The latter value is in good agreement with values reported from ARPES measurements\cite{Zhou2006} and is higher than values obtained from first principle calculations\cite{PhysRevB.43.4579}, but is significantly lower than values reported by several other authors from magneto-optical measurements.\cite{DRESSELHAUS1966465,Toy77,Chung2002} Because all other band structure parameter determined here scale with $\gamma_0$, and in order to facilitate comparison the literature, we use a literature value of $\gamma_0=3.12$~eV ($v_f=1.01\times 10^6$~m/s).\cite{Chung2002}

From $\frac{1}{m^*}=\frac{v_f^2}{\gamma_1}$ (Eqn.~\ref{eqn:effective-masses_b}) we obtain the first inter-layer hopping parameter $\gamma_1=(0.395\pm 0.006)$eV, which is in excellent agreement with literature values.\cite{doi:10.1143/JPSJ.47.199,DRESSELHAUS1966465,PhysRev.134.A453,PhysRevB.19.4224} From $\frac{3\eta-\frac{\gamma_4}{\gamma_0}}{\gamma_1}=(0.420\pm 0.005)$eV$^{-1}$ we determine the second inter-layer hopping parameter $\gamma_4=(0.10\pm 0.02)$eV, which is also in good agreement with the literature.\cite{PhysRevB.21.827,PhysRevB.19.4224} With limited sensitivity to the H-point transitions $\Delta$ cannot be determined. Similarly $\gamma_2$ and $\gamma_5$ cannot be determined independently by Landau level spectroscopy only (they appear as sums in Eqns.~\ref{eqn:solution}-\ref{eqn:energy_splitting}). Only the term $\gamma_2-\gamma_5-\Delta /2=(0.063\pm 0.013)$eV is evaluated here, corroborating results from Refs.~\onlinecite{doi:10.1080/00018730110113644,Chung2002,PhysRevB.46.4531} and references therein. Note that values of the band structure parameter $\gamma_1,\gamma_4$ and $\gamma_2-\gamma_5-\Delta/2$ justify the assumptions made in the derivation in the appendix. The results for the band structure parameter are summarized in Tab.~\ref{tab:BestModelSWMGraphite}.


\subsubsection{Graphene}
Free charge carriers in the epitaxial graphene layer are modeled with a two channel model,\cite{HofmannAPL98_2011} where the effective mass is coupled to the carrier concentration.\cite{NovoselovN438_2005} The behavior of the on-diagonal block elements of $\mathbf{M}^+$ (not shown) and $\mathbf{M}_0$ (not shown) in the reststrahlen band ($100-120$~meV, pink bar in Figs.~\hyperref[fig:graphite+,-]{\ref*{fig:graphite+,-}b}, \hyperref[fig:graphite-graphene_field_dependence]{\ref*{fig:graphite-graphene_field_dependence}c} and \hyperref[fig:graphite-graphene_field_dependence]{\ref*{fig:graphite-graphene_field_dependence}d}) indicates a high carrier concentration and a low mobility. However, the off-diagonal block elements of the OHE data indicate a high mobility and a low carrier concentration. An explanation may be found in the high number of graphene sheets present in the sample ($10-20$), where the doping decreases and the mobility increases with the distance from the polar interface of the C-face 6\textit{H}-SiC substrate. The best-match model parameters are similar to electrical Hall measurements and are in good agreement with the literature.\cite{TedescoAPL95_2009,HofmannAPL98_2011} The TO-LO phonon resonance in the 6\textit{H}-SiC substrate yields for $\varepsilon_{\perp}=\varepsilon_{x}=\varepsilon_{y}$ the best-match-model parameters 
\mbox{$\omega_{\text{TO},\perp}=(99.10\pm 0.01)$meV}, \mbox{$\omega_{\text{LO},\perp}=(120.60\pm 0.01)$meV}, \mbox{$\gamma_{\text{TO},\perp}=(0.15\pm 0.02)$meV}, \mbox{$\gamma_{\text{LO},\perp}=(0.58\pm 0.02)$meV} and \mbox{$\varepsilon_{\infty,\perp}=(6\pm 0.1)$}. For \mbox{$\varepsilon_{\parallel}=\varepsilon_{z}$} the best-match-model parameters are \mbox{$\omega_{\text{TO},\parallel}=(99.0\pm 0.5)$meV}, $\omega_{\text{LO},\parallel}=(119.89\pm 0.01)$meV, $\gamma_{\text{TO},\parallel}=0.06$~meV, $\gamma_{\text{LO},\parallel}=(0.33\pm 0.01)$meV and $\varepsilon_{\infty,\parallel}=(5.8\pm 0.2)$, where the parameter $\gamma_{\text{TO},\parallel}$
 was not varied.

Symbols in Fig.~\hyperref[fig:Graphene_transition_energies_0]{\ref*{fig:Graphene_transition_energies_0}d} and \hyperref[fig:Graphene_transition_energies_0]{e} display best-match-model parameter from OHE data analysis using the small energy splitting approximation (Eqn.~\ref{eqn:landau_averaged}) for the energy parameters of the $N^{\text{th}}$ inter-Landau-level transition associated with single and multilayer graphene, respectively. Lines represent calculated data using best-match-model parameters for transition energies of the small energy splitting approximation $E^{\stext{SLG}}_{N}$ and $E^{\stext{m-BLG}}_{N}$ versus the applied magnetic field according to Eqns.~\ref{eqn:SLG_LL} and \ref{eqn:LLNLayer}, respectively. From the polarization mode conserving inter-Landau-level transitions $E^{\stext{SLG}}_{N}$ with $N=1\dots 16$ we determine the best-match-model Fermi velocity parameter as \mbox{$v_f=(1.01\pm 0.01)\times 10^6$m/s}, corresponding to \mbox{$\gamma_0=(3.12\pm 0.03)$eV}, which is in very good agreement with Refs.~\onlinecite{OrlitaPRB83_2011, SadowskiPRL97_2006, OrlitaPRL102_2009, HenriksenPRL100_2008, OrlitaPRL107_2011, OrlitaPRL101_2008}.

The polarization mode mixing inter-Landau-level transitions $E^{\stext{m-BLG}}_{N}$ are identified to originate from bi-, tri-, and quad-layer graphene, with $N^{(2)}=3\dots 9$, $N^{(3)}=3,5,6,7,9,10$ and $N^{(4)}=4,5,9,10$, respectively. Using the Fermi velocity from the polarization mode conserving transitions, the best-match-model parameters for the inter-layer coupling constant $\gamma_1$ for bi-, tri-, and quad-layer graphene are $\gamma_1^{(2)}=(0.387\pm 0.001)$eV, $\gamma_1^{(3)}=(0.385\pm 0.001)$eV and $\gamma_1^{(4)}=(0.404\pm 0.003)$eV, respectively, corroborating experimental result for bi-\cite{PhysRevB.76.201401,PhysRevLett.102.037403,PhysRevB.80.165406,PhysRevB.78.235408,OrlitaPRL107_2011} and tri-layer\cite{Taychatanapat2011} graphene. Note that the parameter $\gamma_1^{(4)}$ is deviating from the corresponding values for bi- and tri-layer graphene due to an imperfect match for the $N=4,5$ transitions when compared to the $N=9,10$ transitions. A possible explanation is that the transition energies for $N=4,5$ (with a low quantum number $N\leq 5$) might be, similar to graphite,\cite{PhysRev.140.A401,doi:10.1143/JPSJ.40.761} affected by the trigonal warping due to $\gamma_3$, which is neglected in the derivation of Eqn.~\ref{eqn:LLNLayer}.

For the highest magnetic field $B_{\perp}=8/\sqrt{2}$~T and the two bi-layer inter-Landau-level transitions with the highest signal-to-noise ratio in the on- and off-diagonal block elements in $\delta\mathbf{M}^{+}$  \mbox{($N=4,5$)}, we apply the circular polarization transition model (Eqn.~\ref{eqn:landau}). Similar to the case of graphite, we set $A_{\pm,N}=A_{N}$ and $\gamma_{\pm,N}=\gamma_{N}$ ($N\geq 2$) for the amplitude and broadening parameters, respectively. The energy splitting between the transition energies for right- and left-handed circular polarized light is set constant \mbox{$\delta E^{\stext{N-BLG}}_{N}=\delta E_0$}. For the two transitions analyzed, we determine \mbox{$\delta E_0=(5.8\pm 1.3)$meV}, which is, within the error margin, identical to the corresponding value in graphite.

According to Eqns.~\ref{eqn:solution}--\ref{eqn:energy_splitting}, the \mbox{$\Gamma$-dependent} effective mass is $m^{*}_{\stext{m-BLG}}=\Gamma\frac{\gamma_1}{2v_{F}^2}$, in accordance with Ref.~\onlinecite{PhysRevLett.96.086805,Novoselov2006,KoshinoPRB77_2008,0034-4885-76-5-056503}, and the energy splitting between right- and left-handed polarized light can be approximated as $\delta E^{\stext{m-BLG}}_{N}\approx 8\eta\hbar\frac{e B}{m^*}$ with $\eta=\frac{\Gamma}{2}\frac{\gamma_4}{\gamma_0}$ (where $\gamma_2=\gamma_5=0$ for bi-layer graphene and $\Delta=0$). Within this approach, using $\Gamma=1$ ($\kappa=\pi/3$)\cite{KoshinoPRB77_2008} for bi-layer graphene, we find \mbox{$m^{*}_{\stext{m-BLG}}=(0.0338\pm 0.0001)m_e$} and therefore $\gamma_4^{(2)}=(0.23\pm 0.06)$eV, which is in good agreement with Refs.~\onlinecite{PhysRevB.78.235408,PhysRevB.84.085408,PhysRevB.80.165406,PhysRevB.76.201401}. The results for the band structure parameters are summarized in Tab.~\ref{tab:BestModelSWMGraphite}.

\section{Conclusions}
We derived and applied a dielectric function tensor model for description of inter-Landau-level transitions and their polarization selection rules, which allowed us to determine tight-binding Slonczewski-Weiss-McClure model parameters from best-match-model calculation of optical Hall effect data. We employed this model for highly oriented pyrolytic graphite and for multilayer graphene on 6\textit{H}-SiC. We thereby studied similarities and differences between the band structures of two- and three-dimensional layered materials with hexagonal crystal symmetries. We used mid-infrared reflection-type optical Hall effect measurements to investigate inter-Landau-level transitions at sample temperatures of $T=1.5$~K and magnetic fields up to $B=8$~T. From the magnetic field dependence we identified H- and K-point inter-Landau-level transitions in graphite and decoupled single-layer and coupled bi-, tri-, and quad-layer inter-Landau-level transition in multilayer graphene. The polarization selection rules were determined as polarization state conserving for H-point transitions in graphite and decoupled single-layer transitions in graphene, while K-point and coupled multilayer transitions were found to have polarization state mixing properties. The consequences from polarization selection rules for the symmetry-properties of the corresponding dielectric tensors were discussed.

\acknowledgments
The authors would like to acknowledge financial support from the Swedish Research Council (VR Contract Nos. 2013–5580 and 2016-00889), the Swedish Governmental Agency for Innovation Systems (VINNOVA) under the VINNMER international qualification program Grant No. 2011-03486, the Swedish foundation for strategic research (SSF) under Grant Nos. FFL12-0181 and RIF14-055, and the Swedish Government Strategic Research Area in Materials Science on Functional Materials at Link€oping University (Faculty Grant SFO Mat LiU No. 2009 00971). This work was supported in part by the National Science Foundation (NSF) through the Center for Nanohybrid Functional Materials (EPS-1004094), the Nebraska Materials Research Science and Engineering Center (DMR 1420645), and Award No. EAR 1521428. The authors further acknowledge financial support by the J.A. Woollam Co., Inc., and the J. A. Woollam Foundation. We thank Professor Pablo Esquinazi, University of Leipzig, Germany, and Professor Kurt Gaskill, U.S. Naval Research Laboratory, Washington, D.C, for providing the graphite and epitaxial graphene samples, respectively.

\appendix

\section{The Slonczewski-Weiss-McClure~(SWC) model}\label{appendix:SWC-model}
Within the SWC model the Hamiltonian of AB-stacked graphene sheets can be written as\cite{PhysRev.109.272,PhysRev.108.612}

\begin{equation} 
	\mathbf{H}
	=
	\begin{pmatrix}
			E_1                & 0                 & -k_{_+} \zeta_{_-} &  k_{_-} \zeta_{_-}  \\
			0                  & E_2               & -k_{_+} \zeta_{_+} & -k_{_-} \zeta_{_+}  \\
			 k_{_-} \zeta_{_-} & k_{_-} \zeta_{_+} & E_3                & -k_{_+} \zeta_{3}    \\
			-k_{_+} \zeta_{_-} & k_{_+} \zeta_{_+} & k_{_-} \zeta_{3}   & E_3
	\end{pmatrix}\;,
	\label{eqn:Hamiltonian-WSC}
\end{equation}

with

\begin{subequations}
	\begin{align}
		E_1 & = \frac{1}{2}\gamma_5\Gamma^2 + \gamma_1\Gamma + \Delta \;, \\
		E_2 & = \frac{1}{2}\gamma_5\Gamma^2 - \gamma_1\Gamma + \Delta \;, \\
		E_3 & = \frac{1}{2}\gamma_2\Gamma^2 \;, \\
		\Gamma & = 2 \cos(\frac{k_z c_0}{2})= 2 \cos \kappa \;, \\
		k_{\pm} & = \frac{1}{2}(k_x \pm \text{i} k_y)\;, \\
		\zeta_{\pm} &= \text{i} \hbar v_F (\Gamma \frac{\gamma_4}{\gamma_0}\pm1) \;, \\
		\zeta_{3} &= \text{i} \hbar v_F \Gamma \frac{\gamma_3}{\gamma_0}\sqrt{2} \;, \\
		v_F &= \frac{\sqrt{3}a_0}{2\hbar} \gamma_0  \;,
		\label{eqn:definitions-WSC}
  \end{align}
\end{subequations}

where $a_0$ and $c_0$ denote the lattice parameters, and $\mathbf{k}=\{ k_x,k_y,k_z \}$ is the wave vector.

For a magnetic field along the z-axis, neglecting $\gamma_3$, which leads to a trigonal wrapping of the band structure and affects mostly Landau levels with small quantum numbers,\cite{PhysRev.140.A401,doi:10.1143/JPSJ.40.761} and using eigenfunctions of a simple harmonic oscillator\cite{doi:10.1143/JPSJ.17.808}, Eqn.~\ref{eqn:Hamiltonian-WSC} can be reduced to an algebraic eigenvalue problem

\begin{widetext}
\begin{equation} 
	\mathbf{H}
	=
	\begin{pmatrix}
			E_1                       & 0                        & -\zeta_{_-} \sqrt{ns} &  \zeta_{_-} \sqrt{(n+1)s}  \\
			0                         & E_2                      & -\zeta_{_+} \sqrt{ns} & -\zeta_{_+} \sqrt{(n+1)s}  \\
			 \zeta_{_-} \sqrt{ns}     & \zeta_{_+} \sqrt{ns}     & E_3                   & 0                   \\
			-\zeta_{_-} \sqrt{(n+1)s} & \zeta_{_+} \sqrt{(n+1)s} & 0                     & E_3
	\end{pmatrix}\;,
	\label{eqn:Hamiltonian-WSC_reduced}
\end{equation}
\end{widetext}
with 
$s=\frac{e B}{\hbar}$ and the quantum number $n$. The characteristic equation then reads
\begin{equation}
\begin{split}
\left(
	\left(
		 E_1 - E
	\right)
	\left(
		E_3 - E
	\right)
	+
	\zeta_{_-}^2 s
	\left(
		2n + 1 
	\right)
\right)& \\
\times
\left(
	\left(
		E_2 - E
	\right)
	\left(
		E_3 - E
	\right)
	+
	\zeta_{_+}^2 s
	\left(
		2n + 1 
	\right)
\right)& \\
- s^2 \zeta_{_-}^2 \zeta_{_+}^2
&
=0\;.
	\label{eqn:characteristic}
\end{split}
\end{equation}
The last term represents the effect of the magnetic field to mix the states of different bands\cite{doi:10.1143/JPSJ.17.808} and is a second order term in $B$.

Koshino and Ando showed,\cite{KoshinoPRB77_2008} while neglecting all tight binding parameters except $\gamma_0$ and $\gamma_1$, that for $m$-layer graphene inter-Landau-level transitions occur at equally spaced positions in \mbox{$k_z$-space} between the K- and H-point with \mbox{$\Gamma_m=2\cos\kappa_m$} and \mbox{$\kappa_m=\frac{k_z c_0}{2}= l\frac{\pi}{m+1}$}, where \mbox{$l=\left\{1,\dots,m-3,m-1\right\}$} for even $m$ and \mbox{$l=\left\{0,\dots,m-3,m-1\right\}$} for odd $m$. Therefore, inter-Landau-level transitions with $\Gamma=0$ occur in single-layer graphene ($m=1$) and multilayer graphene with an odd layer number ($m=3,5,7\dots$).

Transferring this concept to graphite \mbox{($m\rightarrow \infty$)}, inter-Landau-level transitions can occur for every value of $k_z$ between the K- and H-point. However, when applying Fermi's golden rule, it can be shown\cite{PhysRev.140.A401,doi:10.1143/JPSJ.17.808} that the joint density of states (between conduction band and valence band) has singularities for $\Gamma=2$ and $\Gamma=0$, leading to the occurrence of inter-Landau-level transitions in graphite at the two high symmetry points of the Brillouin zone, the H- and K-point, only.
 
For inter-Landau-level transitions with $\Gamma=0$ (single-layer graphene, multilayer graphene with odd layer number and the H-point in graphite), Eqn.~\ref{eqn:characteristic} is highly simplified and the Landau level energies are
\begin{equation}
\begin{split}
	E^{\stext{H}\pm}_{1} &=\frac{\Delta}{2} \pm \sqrt{\frac{B}{b_0} \gamma_0^2 n    + \frac{\Delta}{4}^2}\;, \\
	E^{\stext{H}\pm}_{2} &=\frac{\Delta}{2} \pm \sqrt{\frac{B}{b_0} \gamma_0^2 (n+1) + \frac{\Delta}{4}^2}\;,
\end{split}
\end{equation}
with the characteristic magnetic field $b_0=\frac{2\hbar}{3 a_0^2 e}$. The two solutions are identical but shifted by 1 in $n$, leading to a double degeneracy for all levels except $E^{\stext{H}\pm}_{1}(n=0)$. 

For all other transitions ($\Gamma\neq 0$) the field mixing term in Eqn.~\ref{eqn:characteristic} is neglected. Keeping the order of the bands in mind, i.e., \mbox{$E_{1}-E_{3}>0$} and \mbox{$E_{2}-E_{3}<0$}, the Landau level energies can be written as
\begin{equation}
\begin{split}
	E^{\pm}_{1} &\hspace{-1pt}=\hspace{-1pt}E_{3}\hspace{-1pt}+\hspace{-1pt}\frac{E_{1}\hspace{-1pt}-\hspace{-1pt}E_{3}}{2}\hspace{-3pt}\left(\hspace{-3pt} 1 \hspace{-2pt} \pm 
    \hspace{-3pt}\sqrt{\hspace{-2pt} 1 \hspace{-2pt} - \hspace{-2pt} \frac{8s\zeta_{_-}^2(n+\frac{1}{2})}{\left(E_{1} - E_{3}\right)^2}}\right)\;, \\
	E^{\pm}_{2} &\hspace{-1pt}=\hspace{-1pt}E_{3}\hspace{-1pt}+\hspace{-1pt}\frac{E_{2}\hspace{-1pt}-\hspace{-1pt}E_{3}}{2}\hspace{-3pt}\left(\hspace{-3pt} 1 \hspace{-2pt} \mp 
    \hspace{-3pt}\sqrt{\hspace{-2pt} 1 \hspace{-2pt} - \hspace{-2pt} \frac{8s\zeta_{_+}^2(n+\frac{1}{2})}{\left(E_{1} - E_{3}\right)^2}}\right).
\label{eqn:solution}
\end{split}
\end{equation}

The two branches close to $E_{3}$, relevant for the inter-Landau-level transitions observed in this paper, are $E^{+}_{1}$ and $E^{-}_{2}$ and will be further called conduction (c) and valence (v) band, respectively. The second term under the square root is rewritten using \mbox{$\gamma_0\gg\gamma_4$} and \mbox{$\gamma_1\gg\gamma_2-\gamma_5-\frac{\Delta}{2}$} and therefore neglecting all higher order terms in $\frac{\gamma_4}{\gamma_0}$ and $\frac{2\gamma_2-2\gamma_5-\Delta}{2\gamma_1}$ and their products\cite{PhysRevB.21.827} 
\begin{subequations}
	\begin{align}
		\frac{2s\zeta_{_{\mp}}^2(n+\frac{1}{2})}{E_{\stackrel{1}{\text{\tiny{\textit{2}}}}} - E_{3}}
		& \approx \pm \hbar\frac{e B}{m^*}\left(1\pm 4\eta\right)=\pm \hbar\frac{e B}{m_{\stackrel{c}{\text{\tiny{\textit{v}}}}}}
				\label{eqn:effective-masses_a}
\;, \\
		\frac{1}{m^*} & = \frac{v_F^2}{\gamma_1}\frac{2}{\Gamma} = \frac{3a_0^2\gamma_0^2}{2\Gamma\hbar^2\gamma_1}
				\label{eqn:effective-masses_b}
\;, \\
		\eta &= \frac{\Gamma}{2}\frac{\gamma_4}{\gamma_0} - \frac{\frac{\Gamma^2}{2}\gamma_2-\frac{\Gamma^2}{2}\gamma_5-\Delta}{8\gamma_1}\;.
		\label{eqn:effective-masses_c}
  \end{align}
\end{subequations}

Expanding Eqn.~\ref{eqn:solution} up to the second order in $B$ it is rewritten as
\begin{equation}
\begin{split}
	E_{\stackrel{c}{\text{\tiny{\textit{v}}}}} = E_{3} \pm \hbar \frac{e B}{m^*}& \left(n+\frac{1}{2}\right)\left(1\pm 4\eta\right)\\
  - \left(\hbar \frac{e B}{m^*}\right)^2
	& \left(n+\frac{1}{2}\right)^2
	\frac{8(\frac{1}{\Gamma}+1)\eta-4\frac{\gamma_4}{\gamma_0}\pm 1}{\Gamma\gamma_1}\;,
\end{split}
\end{equation}
where we set
\begin{equation}
		\frac{1}{E_{\stackrel{1}{\text{\tiny{\textit{2}}}}} - E_{3}}\left(1\pm 4\eta\right)^2
\approx
		-\frac{8(\frac{1}{\Gamma}+1)\eta-4\frac{\gamma_4}{\gamma_0}\pm 1}{\Gamma\gamma_1}\;,
\end{equation}
using the same approximations as for Eqns.~\ref{eqn:effective-masses_a}-\ref{eqn:effective-masses_c}.

Inter-Landau-level transition energies are determined by Eqn.\ref{eqn:transition_energies} and obey, neglecting trigonal warping,\cite{doi:10.1143/JPSJ.40.761,PhysRev.140.A401,doi:10.1143/JPSJ.17.808} the selection rules \mbox{$n'-n=1$}, with $n=\{0,1,2\dots\}$. In the following (+) and (-) stand for right- and left-handed circular polarized inter-Landau-level transitions, respectively. Using the quantum number $2N=n+n'+1$, with $N=\{1,2,3\dots\}$ the transition energies are
\begin{equation}
\begin{split}
	E_{N,\pm} & = \hbar\frac{e B}{m^*}\left(2N\pm 4\eta \right)\\
	- & \left(\hspace{-1pt} \hbar\frac{e B}{m^*}\hspace{-1pt}\right)^2\hspace{-3pt}
	\left(
		\frac{4(\frac{1}{\Gamma}\hspace{-2pt}+\hspace{-2pt}1)\eta\hspace{-2pt}-\hspace{-2pt}2\frac{\gamma_4}{\gamma_0}}{\Gamma\gamma_1}\hspace{-1pt}\left(4N^2\hspace{-2pt}+\hspace{-2pt}1\right)\hspace{-2pt}\pm\hspace{-2pt}\frac{2N}{\Gamma\gamma_1}
	\right).
\end{split}
\end{equation}
Finally, we determine the transition energies $E_{N}$ and the energy splitting $\delta E_{N}$ for the small energy splitting approximation
\begin{subequations}
\begin{align}
\begin{split}
		E_{N} = & \hbar\frac{e B}{m^*}2 N\\
		& -
		\left(\hspace{-2pt}\hbar\frac{e B}{m^*}\hspace{-2pt}\right)^2
		\hspace{-5pt}\left(4 N^2\hspace{-2pt}+\hspace{-2pt}1\right)
		\hspace{-2pt}\frac{4(\frac{1}{\Gamma}\hspace{-2pt}+\hspace{-2pt}1)\eta\hspace{-2pt}-\hspace{-2pt}2\frac{\gamma_4}{\gamma_0}}{\Gamma\gamma_1} \;,
			\label{eqn:circ_av_model}
\end{split}
			\\
\begin{split}
    \delta E_{N} = & \hbar\frac{e B}{m^*} 8\eta \\
			& +
			\left(\hspace{-2pt}\hbar\frac{e B}{m^*}\hspace{-2pt}\right)^2
			\frac{4N}{\Gamma\gamma_1}\;.
			\label{eqn:energy_splitting}
\end{split}
\end{align}
\end{subequations}
\ifthenelse{\boolean{ext_bbl}}{

}{
\bibliography{../../../CompleteLibrary}}

\begin{thebibliography}{100}
\bibitem{Note_PRB1}
In the following equation the Einstein notation is used, and the covariance and
  contra variance is ignored since all coordinate systems are Cartesian (The
  summation is only executed over pairs of lower indices).
\bibitem{Note_PRB2}
This statement stays correct when effects such as trigonal warping of the band
  structure in graphite are included\cite
  {doi:10.1143/JPSJ.40.761,PhysRev.140.A401,doi:10.1143/JPSJ.17.808}.
\bibitem{Note_PRB3}
Using Fermi's golden rule it can be shown\cite
  {PhysRev.140.A401,doi:10.1143/JPSJ.17.808} that the joint density of states
  (between conduction band and valence band) has singularities at the two high
  symmetry points of the Brillouin zone, the H- and K-point.
\bibitem{Note_PRB4}
Note that all Mueller matrix elements of the GE data discussed in this paper,
  are normalized by the element $M_{11}$, therefore $|M_{ij}|\leq 1$ and
  therefore $M_{11}=1$.
\bibitem{Note_PRB7}
Note, that the individual graphene layers of $m$-layer graphene are optically
  indistinguishable in this sense, since they can be described by the same
  dielectric tensor.
\bibitem{Note_PRB8}
Note that the ellipsometer used here is a rotating analyzer ellipsometer and
  cannot measure the last row and column of the Mueller matrix.
\bibitem{Note_PRB9}
Sub-linear behavior means $B_{\perp }^x$ with $1> x > \protect \frac {1}{2}$,
  and can be observed comparing the line shapes in Fig.~\protect \hyperref
  [fig:Graphene_transition_energies_0]{\ref
  *{fig:Graphene_transition_energies_0}d}~and~\protect \hyperref
  [fig:Graphene_transition_energies_0]{e}.
\bibitem{Note_PRB10}
This includes higher order optical selection rules with $n'\not =n\pm 1$\cite
  {PhysRev.140.A401}, which are neglected here.
\bibitem{Note_PRB11}
Note that the experimental setup does contain linear polarizers only.
\end{thebibliography}
\bibliographystyle{APLstyleNOwebref_2}

\end{document}